\documentclass[useAMS,usenatbib,usegraphicx]{mn2e}
\usepackage{graphicx}

\title[Centrosymmetric molecules as possible carriers of DIBs]
{Centrosymmetric molecules as possible carriers of diffuse interstellar bands
\thanks{Based on observations made with ESO Telescopes at the Paranal Observatory under programme IDs 266.D-5655(A),
67.C-0281(A), 71.C-0513(C), 67.D-0439(A), 082.C-0566(A) and La Silla under programme IDs 078.C-0403(A), 076.C-0164(A) and 073.C-0337(A);
and also observations made with the 1.8m telescope in South Korea and 2m telescope in International Centre for Astronomical and Medico-Ecological Research, Terskol, Russia.
}}
\author[M. Ka{\'z}mierczak et al.]
{M. Ka{\'z}mierczak,$^{1}$
\thanks{e-mail: kazmierczak@astri.uni.torun.pl}
M. R. Schmidt,$^{2}$
\thanks{e-mail: schmidt@ncac.torun.pl}
G. A. Galazutdinov,$^{3}$
\thanks{email: runizag@gmail.com}
F. A. Musaev,$^{4}$
\newauthor 
Y. Betelesky,$^{5}$
\thanks{email: ybialets@eso.org}
J. Kre{\l}owski$^{1}$
\thanks{e-mail: jacek@astri.uni.torun.pl}
\\
\medskip \\
$^1$ Centre for Astronomy, Nicolaus Copernicus University,
Gagarina 11, 87-100 Toru{\'n}, Poland\\
$^2$ Nicolaus Copernicus Astronomical Center, 
Rabia{\'n}ska 8, 87-100 Toru{\'n}, Poland\\
$^3$ Instituto de Astronomia, 
Universidad Catolica del Norte,
Angamos 0610, Antofagasta, Chile\\
$^4$ International Centre for Astronomical 
and Medico-Ecological Research, Terskol, Russia\\
$^5$ European Southern Observatory, 
Karl-Schwarzschild-Strasse 2, 85748 Garching, Germany
}

\date{Accepted...
      Received ...
      in original form ...}

\pagerange{\pageref{firstpage}--\pageref{lastpage}}
\pubyear{2008}

\begin{document}

\maketitle

\label{firstpage}

\begin{abstract}
This paper presents a new data with interstellar $C_2$ (Phillips bands $\rm A^{1}\Pi_u-X^{1}\Sigma^{+}_g$) which were observed with ESO/UVES spectrograph. We determined interstellar column densities and excitation temperatures of $C_2$ for nine Galactic lines. For seven of them $C_2$ have never been observed before, so in that case still small sample of interstellar clouds (26 lines of sight) where a detailed analysis of excitation of $C_2$ was made, increased significantly. This paper is a continuation of previous works where interstellar molecules ($C_2$ and diffuse interstellar bands) were analysed. Since a sample of interstellar clouds with $C_2$ risen we can show that the width and shape of some DIB's profile (6196\,\AA, 5797\,\AA) apparently depend on the gas kinetic and rotational temperature of  $C_2$, being broader for its higher values. There are also DIBs (4964\,\AA, 5850\,\AA) for which that effect does not exist.

\end{abstract}

\maketitle

\section{Introduction}

$C_2$ is the simplest multi-carbon species. As a homonuclear diatomic molecule, has a negligible dipole moment and hence radiative cooling of the excited rotational levels may go only through the slow quadrupole transitions \citep{Dishoeck1982}. The rotational levels are pumped by the galactic interstellar radiation field and excited effectively above the gas kinetic temperature. Lines of the diatomic carbon from a long-lived ground state rotational levels are measurable and can be the sensitive diagnostic probes of conditions in molecular clouds that produce the interstellar absorption lines, in contrast to polar molecules, such as $CH$ or $CN$, where usually only a few absorption lines from the lowest rotational levels are observed. $C_2$ is a useful tool to determine physical conditions (temperatures and densities) in interstellar clouds. 

Moreover, $C_2$ abundances may give information on the chemistry of the intervening clouds, especially on the pathway to the formation of long chain carbon molecules which may be connected with carriers of diffuse interstellar bands (DIBs) \citep{Douglas1977,Maier2006}. Diffuse interstellar bands (DIBs) were first time observed in interstellar medium by \citet{Heger1922}. At present we know 414 DIBs \citep{Hobbs2009}, but none of them has been identified yet in spite of being the subject of much observational and theoretical research.

Useful information allowing identification of diffuse interstellar bands may come from an analysis of their profiles. They can vary \citep{Herbig1975,Westerlund1988,Galazutdinov2002} or have some substructures inside the profiles \citep{Sarre1995} not only because of multi interstellar clouds toward one star causing the Doppler splitting which can likely modify the profiles of DIBs \citep{Herbig1982}. \citet{Kazmierczak2009aa} showed that there is a relation between the profile widths of strong diffuse interstellar band at 6196\,\AA~and the excitation temperatures of $C_2$. The 6196\,\AA~carrier could be a centrosymmetric molecule, whose spectral features become broader as their rotational temperatures increase. 

The goal of this paper is at first to increase sample of interstellar clouds where a detailed analysis of excitation of $C_2$ was made (since that time we had known 26 objects with interstellar $C_2$, 24 lines of sight - see \citep{Sonnentrucker2007} - Table 13 plus 2 new objects in \citep{Kazmierczak2009mnras}).

The second goal of this paper is to analyse other diffuse interstellar bands in comparison to excitation temperatures of $C_2$. We found that some of DIB profiles (6196\,\AA~ \citep{Kazmierczak2009aa}, 5797\,\AA~(this paper)) widths apparently depend on the $C_2$ temperature, being broader for its higher values. There are also DIBs (4964\,\AA, 5850\,\AA) for which that effect does not exist. Although effect is very subtle, the difference is evident.

In Section 2 and 3, we describe $C_2$ and analysed diffuse interstellar bands, respectively. General discussion and summary of our conclusions
are given in Section 4.

\section{$C_2$}
\subsection{The observational data}

Observations of the program objects (Table 1) were made in March 2009, using the high-resolution spectrograph UVES (UV-Visual Echelle Spectrograph) of the VLT fed by the Kueyen telescope of the ESO Paranal Observatory, Chile. That part of work is a continuation of our previous paper \citep{Kazmierczak2009mnras} based on archive UVES spectra. The spectral analysis and following calculation of column densities, rotational temperatures of $C_2$ and modeling gas kinetic temperatures and density of the collisional partners were made in the same way as it was in details described in \citep{Kazmierczak2009mnras}.

All of the selected targets have one $C_2$ Doppler component at the resolution (R$\sim$85\,000) of the observational material. Generally, we cannot exclude the existence of multiple closely spaced components, moreover it is very likely that toward all of the stars there are at least two different interstellar clouds. 
The presence of weak components should not contaminate weak $C_2$ lines, so it is not so big problem to measure dicarbon molecule features, but when we want to make a comparison with other molecules or DIBs we have to be careful. Because of that, not all objects from Table 1 were used to analyse diffuse interstellar bands (e.g. toward HD\,148379 there are more than one dominant Doppler component in CH 4300\,\AA\, and KI 7698\,\AA\, interstellar lines).

Part of data reduction was made with the DECH20T code (Galazutdinov 1992) and with {\it IRAF}\footnote{The Image Reduction and Analysis Facility ({\em IRAF}) is distributed by the National Optical Astronomy Observatories, which is operated by the Association of Universities for Research in Astronomy, Inc. (AURA), under cooperative agreement with the National Science Foundation.} which was also used to spectral analysis.

\begin{table}
\centering
\caption{Basic data for the programme stars}
\label{tab01}
\begin{tabular}{lccc}
\hline
object & name& Sp/L & E(B-V)\\
\hline
HD\,115842 &           & B0.5Ia  & 0.49\\
HD\,136239 & CZ Cir    & B1.5Ia  & 1.11\\
HD\,148379 &           & B2Iab   & 0.71\\
HD\,149757 &$\zeta$ Oph& O9.5V   & 0.28\\
HD\,151932 &           & WN7A    & 0.35\\
HD\,152236 & 	       & B1Iape  & 0.66\\
HD\,154368 & V1074 Sco & O9.5Iab & 0.78\\
HD\,154445 & 	       & B1V     & 0.35\\
HD\,170740 & 	       & B2V     & 0.45\\
\hline
\end{tabular}
\end{table}

\subsection{Results}

We have identified absorption lines of the (1,0), (2,0), (3,0) bands of the $C_2$ Phillips system \mbox{($\rm A^{1}\Pi_u-X^{1}\Sigma^{+}_g$)} (P, Q, R branches in bands (1,0) \mbox{10133 - 10262 {\AA}}, (2,0) \mbox{8750 - 8849 {\AA}}, (3,0) \mbox{7714 - 7793 {\AA}}). The equivalent widths with errors of all measured interstellar lines of $C_2$ towards our program stars are given in Tables 2-4. 
\begin{table*}
{\vbox{\footnotesize
	\tabcolsep=2.5pt
\label{tab02}
\centering
\caption{Observation summary table with equivalent widths (the uncertainties of the $\Delta EWs$ were estimated with the {\it IRAF}; the errors were propagated to the uncertainties of the determined parameters e.g. excitation temperatures, column densities; 
and used in search of the best fit model parameters) [m\AA] of $C_2$ (1,0) Phillips lines toward program stars. In table are also $B(N''=J'')$ - branch identification ($J''$ - low rotational level) and $\lambda$ - wavelength in air in \AA~(see text for the references).
}
\begin{tabular}{@{}ccccccccccc}
\hline
$B(N''=J'')$&HD115842& HD136239  & HD148379   & HD149757  & HD151932  & HD152236  & HD154368 & HD154445 & HD170740 & $\lambda$ [{\AA}]\\
\hline  
 $R(6)$ &     	    & $3.2\pm0.9$&$2.2\pm0.9$ &$2.2\pm0.6$&$2.3\pm1.0$&$1.4\pm0.4$& $4.9\pm0.4$&$2.6\pm0.9$&           & 10133.603 \\
 $R(8)$ &$0.8\pm0.6$& $2.5\pm0.8$&$1.2\pm1.0$ &$1.1\pm0.8$&$0.5\pm0.5$&$1.1\pm0.9$& $3.3\pm0.4$&           &           & 10133.854 \\ 
 $R(4)$ &$1.4\pm0.6$& $4.7\pm0.9$&            &$1.9\pm1.1$&$2.5\pm0.8$&$1.8\pm0.4$&$10.0\pm0.3$&           &$4.9\pm1.6$& 10135.149 \\
 $R(10)$&     	    &            &            &           &           &           & $1.0\pm0.4$&$1.2\pm0.6$&           & 10135.923 \\
 $R(2)$ & 	    & $4.4\pm0.7$&$1.4\pm0.7$ &$4.6\pm1.0$&$4.0\pm0.7$&$2.3\pm0.4$&$11.6\pm0.3$&           &$7.5\pm1.7$& 10138.540 \\
 $R(12)$&     	    & $0.9\pm0.6$&            &$1.0\pm0.9$&$0.8\pm0.8$&           & $1.1\pm0.3$&           &           & 10139.805 \\
 $R(0)$ &$1.6\pm0.6$& $4.2\pm0.7$&            &$1.8\pm1.5$&$4.5\pm1.1$&$3.1\pm0.5$& $8.3\pm0.4$&           &           & 10143.723 \\
 $R(14)$&    	    & $0.6\pm0.6$&            &           &           &           &            &           &           & 10145.505 \\
 $Q(2)$ &$1.8\pm0.5$&$11.1\pm1.1$&$2.4\pm0.6$ &$2.7\pm0.7$&$5.8\pm0.8$&$4.6\pm0.5$&$13.4\pm0.4$&           &$4.2\pm0.4$& 10148.351 \\
 $Q(4)$ &$3.1\pm0.6$& $9.1\pm1.1$&            &$5.1\pm0.9$&$4.7\pm0.7$&$3.3\pm0.4$& $9.3\pm0.4$&$1.8\pm0.9$&$4.6\pm0.4$& 10151.523 \\
 $P(2)$ &$0.7\pm0.6$&            &            &           &           &           & $2.9\pm0.4$&$2.3\pm0.9$&           & 10154.897 \\
 $Q(6)$ &$1.7\pm0.6$& $6.1\pm1.1$&            &$2.6\pm0.7$&$3.6\pm0.9$&$3.4\pm0.4$& $9.3\pm0.4$&$2.3\pm0.9$&$2.4\pm0.6$& 10156.515 \\
 $Q(8)$ &$1.5\pm0.6$&            &$0.9\pm0.8$ &$1.7\pm0.9$&           &$1.4\pm0.4$& $7.5\pm0.4$&           &$3.5\pm0.9$& 10163.323 \\
 $P(4)$ &$0.7\pm0.5$& $2.4\pm0.9$&            &           &           &$1.1\pm0.4$& $3.8\pm0.4$&           &           & 10164.763 \\
 $Q(10)$&	    &            &            &$1.3\pm0.8$&           &$0.9\pm0.5$& $3.6\pm0.4$&           &$1.4\pm0.4$& 10171.963 \\
 $P(6)$ &    	    &            &            &           &           &           & $2.2\pm0.4$&           &           & 10176.252 \\
 $Q(12)$&    	    &            &            &           &           &           & $1.7\pm0.4$&           &$1.5\pm0.4$& 10182.434 \\
 $P(8)$ &$0.6\pm0.5$& $1.8\pm0.8$&            &           &           &           & $2.1\pm0.4$&           &           & 10189.693 \\
 $P(10)$&    	    &            &            &           &           &           & $1.0\pm0.4$&           &           & 10204.998 \\
\hline
\end{tabular}
}}
\end{table*}
\begin{table*}
{\vbox{\footnotesize
	\tabcolsep=2.5pt
\label{tab03}
\centering
\caption{The same as in Table 2 but for the (2,0) Phillips band.
Wavelengths marked with 1 were computed from spectroscopical constants of Douay et al. 1988.}
\begin{tabular}{@{}ccccccccccc}
\hline
$B(N''=J'')$&HD115842& HD136239 & HD148379  & HD149757  & HD151932  & HD152236 & HD154368 & HD154445 & HD170740 & $\lambda$ [{\AA}]\\
\hline  
 $R(6)$ & 	    &$1.6\pm0.3$&           &$0.8\pm0.5$&$2.0\pm0.3$&$0.9\pm0.4$&$2.0\pm0.3$&$0.6\pm0.5$&$0.3\pm0.3$& 8750.847 \\
 $R(8)$ &$0.3\pm0.2$&           &$0.5\pm0.2$&$0.9\pm0.6$&$1.3\pm0.3$&$0.6\pm0.4$&$1.7\pm0.3$&$0.7\pm0.6$&           & 8751.487 \\ 
 $R(4)$ &$1.2\pm0.3$&$1.9\pm0.3$&$0.5\pm0.3$&$1.4\pm0.7$&$2.3\pm0.3$&$1.2\pm0.5$&$3.8\pm0.3$&$0.8\pm0.5$&$0.5\pm0.3$& 8751.684 \\
 $R(10)$&$0.2\pm0.2$&           &$0.3\pm0.3$&           &$0.4\pm0.4$&$0.7\pm0.7$&           &$0.3\pm0.3$&$0.3\pm0.3$& 8753.578 \\
 $R(2)$ &$0.8\pm0.3$&$3.9\pm0.2$&$0.9\pm0.3$&$1.4\pm0.5$&$2.3\pm0.3$&$1.4\pm0.3$&$5.3\pm0.3$&$0.6\pm0.3$&$1.2\pm0.4$& 8753.945$^1$ \\
 $R(12)$&           &           &           &$0.3\pm0.3$&           &$0.2\pm0.2$&           &           &           & 8757.127 \\
 $R(0)$ &$1.0\pm0.3$&$3.1\pm0.4$&$0.6\pm0.3$&$1.1\pm0.7$&$1.8\pm0.3$&$0.9\pm0.3$&$4.4\pm0.3$&$0.7\pm0.4$&$1.1\pm0.9$& 8757.683 \\
 $Q(2)$ &$1.3\pm0.3$&$4.8\pm0.7$&$1.5\pm0.3$&$2.4\pm0.5$&$2.6\pm0.3$&$1.7\pm0.3$&$5.9\pm0.3$&$1.1\pm0.3$&$1.1\pm0.4$& 8761.194 \\
 $Q(4)$ &$1.9\pm0.3$&$4.4\pm0.6$&$1.2\pm0.3$&$2.5\pm0.5$&$3.1\pm0.3$&$1.5\pm0.3$&$5.6\pm0.3$&$0.9\pm0.3$&$1.6\pm0.6$& 8763.751 \\
 $P(2)$ &           &$0.7\pm0.3$&           &           &           &$0.2\pm0.2$&$1.3\pm0.3$&$0.3\pm0.4$&           & 8766.026$^1$ \\
 $Q(6)$ &$0.9\pm0.3$&$2.8\pm0.5$&$1.0\pm0.3$&$1.8\pm0.5$&$2.4\pm0.4$&$1.4\pm0.3$&$4.2\pm0.3$&$0.7\pm0.3$&$1.3\pm0.5$& 8767.759 \\
 $Q(8)$ &$0.8\pm0.3$&$0.8\pm0.5$&           &           &$1.9\pm0.8$&$0.3\pm0.3$&$2.8\pm0.3$&$0.6\pm0.4$&           & 8773.220 \\
 $P(4)$ &           &$0.7\pm0.4$&$0.5\pm0.3$&$0.7\pm0.6$&           &$0.4\pm0.3$&$2.3\pm0.3$&           &           & 8773.422$^1$ \\
 $Q(10)$& 	    &$0.8\pm0.3$&           &$1.0\pm0.7$&$1.2\pm0.3$&$0.5\pm0.4$&$2.0\pm0.4$&$0.6\pm0.5$&$0.7\pm0.5$& 8780.141 \\
 $P(6)$ &  	    &           &$0.4\pm0.3$&           &           &           &$1.6\pm0.4$&           &           & 8782.308 \\
 $Q(12)$&           &$0.6\pm0.3$&$0.3\pm0.3$&$0.6\pm0.4$&$0.7\pm0.6$&           &$1.1\pm0.4$&           &$0.6\pm0.6$& 8788.558 \\
 $P(8)$ &           &           &           &           &           &           &$1.6\pm0.4$&           &           & 8792.649 \\
 $Q(14)$&           &           &           &           &$0.5\pm0.4$&           &$1.4\pm0.4$&           &           & 8798.459 \\
 $Q(16)$&     	    &           &           &           &$0.3\pm0.3$&           &$1.3\pm0.4$&           &           & 8809.842 \\
\hline
\end{tabular}
}}
\end{table*}
\begin{table*}
{\vbox{\footnotesize
	\tabcolsep=2.5pt
\label{tab04}
\centering
\caption{The same as in Table 2 but for the (3,0) Phillips band.}
\begin{tabular}{@{}ccccccccccc}
\hline
$B(N''=J'')$&HD115842& HD136239& HD148379  & HD149757  & HD151932 & HD152236 & HD154368 & HD154445 & HD170740 & $\lambda$ [{\AA}]\\
\hline   
$R(6)$ &$0.5\pm0.4$&           &           &$0.4\pm0.4$&$1.0\pm0.7$&$0.2\pm0.2$&$1.2\pm0.3$&$0.8\pm0.7$&$1.3\pm0.8$& 7714.575 \\
$R(4)$ &$0.9\pm0.5$&$1.0\pm0.5$&           &$0.8\pm0.6$&$0.8\pm0.6$&$0.6\pm0.6$&$1.4\pm0.3$&$0.5\pm0.4$&           & 7714.944 \\ 
$R(8)$ &$0.4\pm0.4$&           &           &           &$0.6\pm0.5$&$0.4\pm0.3$&           &$0.6\pm0.5$&$1.1\pm0.8$& 7715.415 \\
$R(2)$ &           &$1.5\pm0.5$&           &$0.5\pm0.4$&$0.9\pm0.8$&$0.6\pm0.4$&$2.2\pm0.3$&           &$2.4\pm0.4$& 7716.528 \\
$R(10)$&           &           &           &           &$0.4\pm0.4$&           &           &$0.5\pm0.5$&           & 7717.469 \\
$R(0)$ &           &$3.7\pm0.5$&$0.8\pm0.7$&$0.6\pm0.6$&$1.1\pm0.7$&$0.9\pm0.5$&$1.4\pm0.3$&           &$1.5\pm0.4$& 7719.329 \\
$R(12)$&           &           &           &           &$0.5\pm0.4$&           &$0.2\pm0.3$&           &           & 7720.748 \\
$Q(2)$ &$0.9\pm0.7$&$2.8\pm0.5$&           &$0.8\pm0.5$&           &$0.5\pm0.5$&$2.3\pm0.3$&           &           & 7722.095 \\
$Q(4)$ & 	   &$1.1\pm0.6$&$1.0\pm0.7$&$0.6\pm0.5$&$1.2\pm0.7$&$0.5\pm0.4$&$3.1\pm0.3$&           &$0.9\pm0.5$& 7724.219 \\
$P(2)$ &           &$1.4\pm0.6$&$0.5\pm0.5$&           &           &           &           &           &           & 7725.819 \\ 
$Q(6)$ &$0.4\pm0.3$&$1.1\pm0.5$&           &           &           &$0.3\pm0.3$&$1.6\pm0.4$&$0.4\pm0.4$&           & 7727.557 \\
$P(4)$ &$0.3\pm0.3$&$0.3\pm0.3$&           &$0.5\pm0.5$&           &           &$0.9\pm0.4$&$0.6\pm0.5$&           & 7731.663 \\
$Q(8)$ &$0.5\pm0.4$&           &           &$0.5\pm0.5$&           &           &$1.1\pm0.4$&$0.4\pm0.4$&$0.3\pm0.3$& 7732.117 \\
\hline
\end{tabular}
}}
\end{table*}

For the optically thin case (when the absorption lines are on the linear part of the curve of growth) column density of a rotational level J'' can be derived from the equivalent width $W_{\lambda}$\,[m\AA] of the single absorption line using the relationship \citep{Frisch1972}
\begin{equation}
N_{col} = 1.13 \times 10^{17} {\frac {W_{\lambda}}{f_{ij} \lambda^2}} ~,
\end{equation}
where $\lambda$ is the wavelength in [\AA], $f_{ij}$ is the absorption oscillator strength.
$C_2$ lines are mostly lying on the linear part of the curve of growth.
In this work, only a few lines of HD\,149757 or HD\,154368 are optically thick
and curve of growth method was applied for the derivation of column densities. 
The turbulent velocity was serched through the minimalization of the dispersion of column densities for each level.
We checked various values of the velocity dispersion parameter
\mbox{($b=\,0;\,0.3;\,0.5;\,0.7;\,1;\,1.5\,\rmn{km}\,\rmn{s}^{-1}$)} and 
\mbox{0.5\,$\rmn{km}\,\rmn{s}^{-1}$} was found to give the lowest dispersion.
This value was consequently applied to all of the program stars. It is consistent with the value derived
in other studies of molecular absorptions (e.g. \cite{Gredel1991}; \cite{Crawford1997}).

The energies of the lower rotational level were determined using molecular constants of \cite{Marenin1970}. The wavelengths
are generally determined from laboratory wave numbers of \citet{Chauville1977} and \citet{BallikRamsay1963}
converted to air wavelengths using Edlen's formula following \citet{Morton1991}.
Wavelengths of three lines R(2), P(2) and P(4) of the (2,0) band, absent in \citep{Chauville1977}, were computed with  \citet{Douay1988} spectroscopic constants. According to \citet{Douay1988} the line positions calculated with their constants should be more accurate than the previous measurements.
The oscillator strengths correspond to vibrational oscillator strengths
$f_{10} = 2.38 \times 10^{-3} $, $f_{20} = 1.44 \times 10^{-3}$, 
$f_{30} = 6.67 \times 10^{-4}$.
The oscillator strengths for individual transitions were computed  according to description in \citet{Bakker1996} using their code MOLLEY. Vibrational oscillator strengths were taken from \cite{Langhoff1990} for (1,0) and (2,0) and from \citet{Bakker1997} (citing Langhoff) for (3,0). Source of band origins were \citet{Chauville1977} and \cite{BallikRamsay1963b}.

The resulting total $C_2$ column densities
are shown in Table 5 together with rotational temperatures of $C_2$ and three parameters estimated from the model (gas kinetic temperature, collisional partners density and column density of the $J''=2$ level).

\begin{table*}
\centering
\caption{Summary of the observational data for $C_2$, $T_{02}$ ($T_{04}$, $T_{06}$) - rotational temperature calculated from the two (three, four) lowest rotational levels, $N_{col}$ - total column densities and the results of the model: $T_{kin}$ - gas kinetic temperature, $n_c$ - the effective density of collision partners ($n_c = n_H + n_{H_2}$) and $N_{col}(J''=2)$ - column density derived from $J''=2$~(for more details - see Kazmierczak et al. 2009).}
\label{tab05}
\begin{tabular}{llllc|llc}
\hline
object & $T_{02} [K]$ & $T_{04} [K]$ & $T_{06} [K]$ & $N_{col}
[10^{12} {cm}^{-2}]$ & $T_{kin} [K]$&$n_c [{cm}^{-3}]$& $N_{col}(J''=2) [10^{12} {cm}^{-2}]$\\
\hline
HD\,115842 & $22\pm9$ & $86\pm31$ & $71\pm13$ & $10\pm2$ & $53\pm30$ & $242\pm217$ & $2.5\pm0.3$\\
HD\,136239 & $41\pm12$& $42\pm4$  & $58\pm4$  & $32\pm2$ & $29\pm5$  & $226\pm30$  & $9.4\pm0.4$\\
HD\,148379 & $52\pm91$& $56\pm22$ & $83\pm21$ & $11\pm2$ & $38\pm27$ & $152\pm77$  & $2.4\pm0.3$\\
HD\,149757 & $42\pm42$& $99\pm47$ & $79\pm14$ & $18\pm2$ & $43\pm19$ & $177\pm55$  & $3.9\pm0.4$\\
HD\,151932 & $28\pm7$ & $55\pm7$  & $75\pm7$  & $26\pm2$ & $33\pm8$  & $163\pm20$  & $5.8\pm0.3$\\ 
HD\,152236 & $26\pm8$ & $43\pm5$  & $72\pm7$  & $13\pm2$ & $27\pm7$  & $176\pm24$  & $3.6\pm0.2$\\
HD\,154368 & $39\pm5$ & $47\pm2$  & $58\pm2$  & $57\pm2$ & $30\pm2$  & $204\pm8$   & $14.5\pm0.2$\\
HD\,154445 & $33\pm33$& $48\pm21$ & $89\pm27$ & $10\pm2$ & $18\pm18$ & $101\pm17$  & $2.1\pm0.4$\\
HD\,170740 & $14\pm5$ & $54\pm10$ & $55\pm7$  & $18\pm2$ & $21\pm9$  & $161\pm16$  & $4.2\pm0.3$\\
\hline
\end{tabular}
\end{table*}

Following \cite{Dishoeck1982} we present derived column densities in form of rotational diagrams (\mbox{Fig. 1}) where weighted relative column densities \mbox{$-ln[5N_{col}(J'')/(2J''+1)N_{col}(2)]$} are plotted versus energy of lower level E''/k (where E'' is the energy of the rotational level J'' and k is the Boltzmann constant). The slope of a straight line on this diagram is nicely connected to the excitation temperature, $a=-1 / T_{exc}$. It is well known from previous works (e.g. \citet{Dishoeck1982}) that populations of all rotational levels cannot be characterised by a single rotational temperature. The lowest J'' levels are described by the lower excitation temperature than higher levels. Such behaviour of the rotational levels was nicely described in the model of excitation of $C_2$ by \cite{Dishoeck1982}.
For the interpretation of the rotational diagrams we have constructed a grid of models based on the radiative excitation model of \citet{Dishoeck1982}. Three independent parameters: gas kinetic temperature ($T_{kin}$), collisional partners density $n_c$, and column density of the $J''=2$ level were estimated simultaneously for the set of observed column densities and the best fitted models are presented on the rotational diagrams.
\begin{figure*}
\centerline{
\hbox{
\includegraphics[width=0.31\textwidth,angle=180]{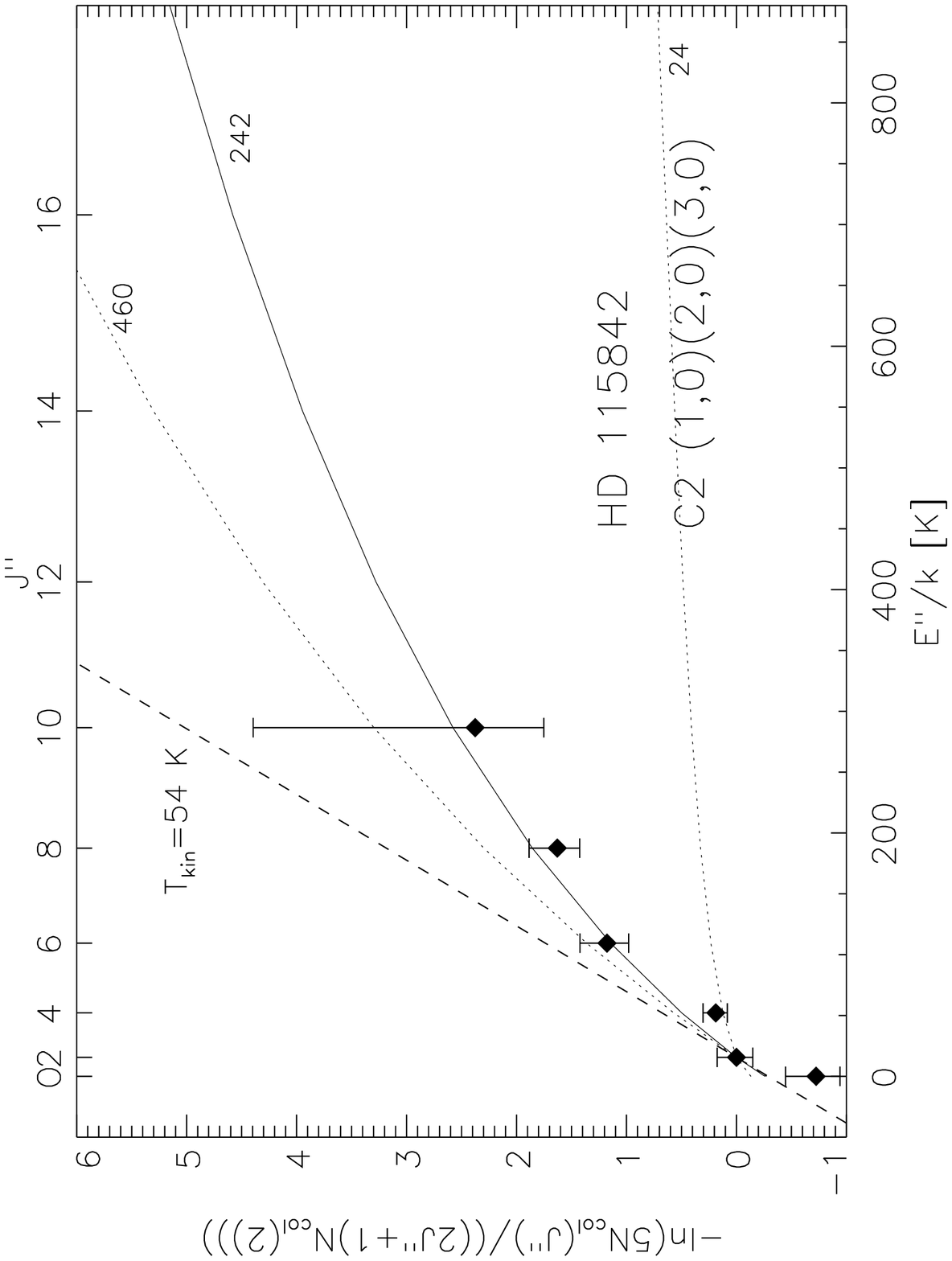}
\includegraphics[width=0.31\textwidth,angle=180]{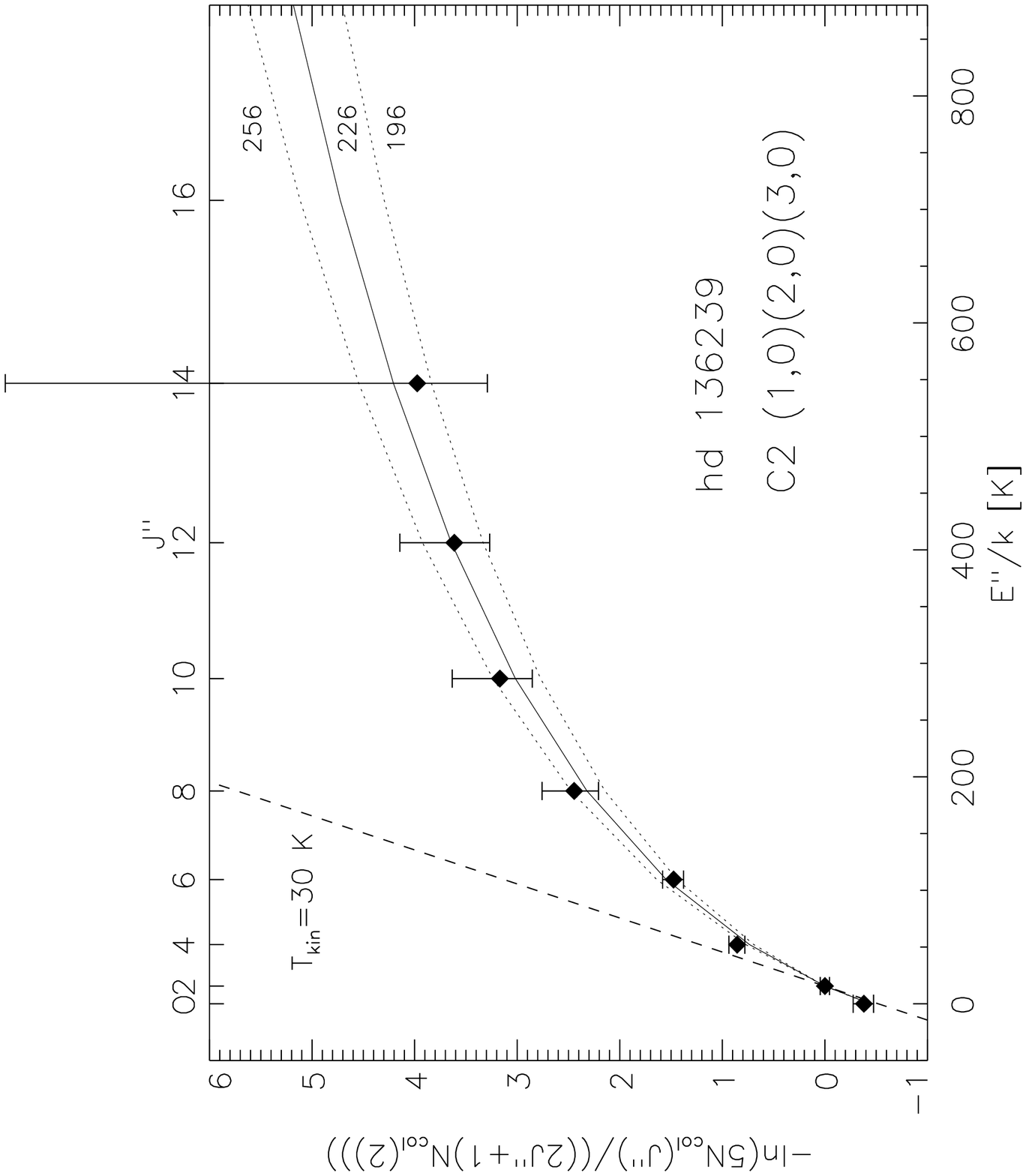}
\includegraphics[width=0.31\textwidth,angle=180]{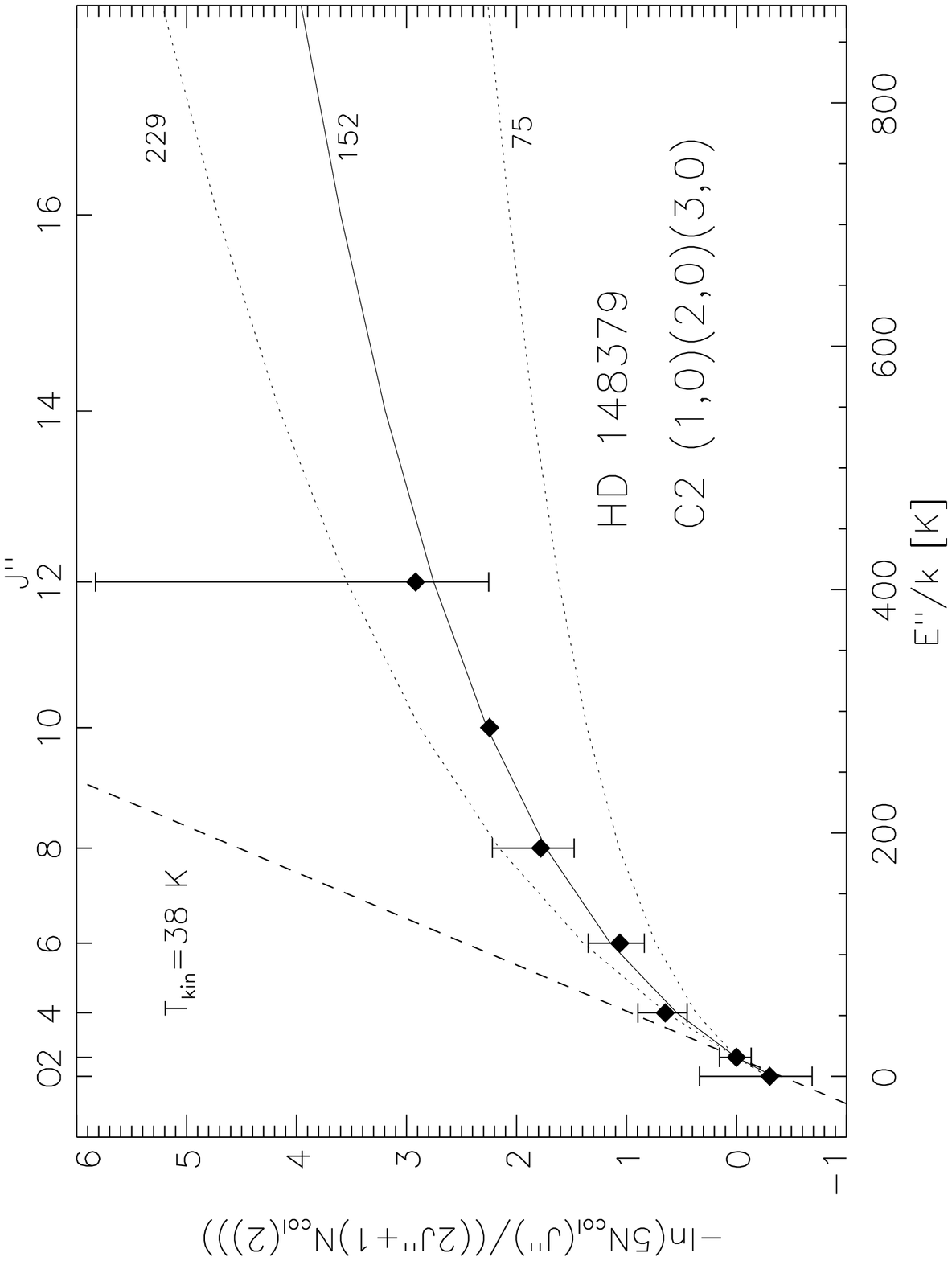}}}
\centerline{
\hbox{
\includegraphics[width=0.31\textwidth,angle=180]{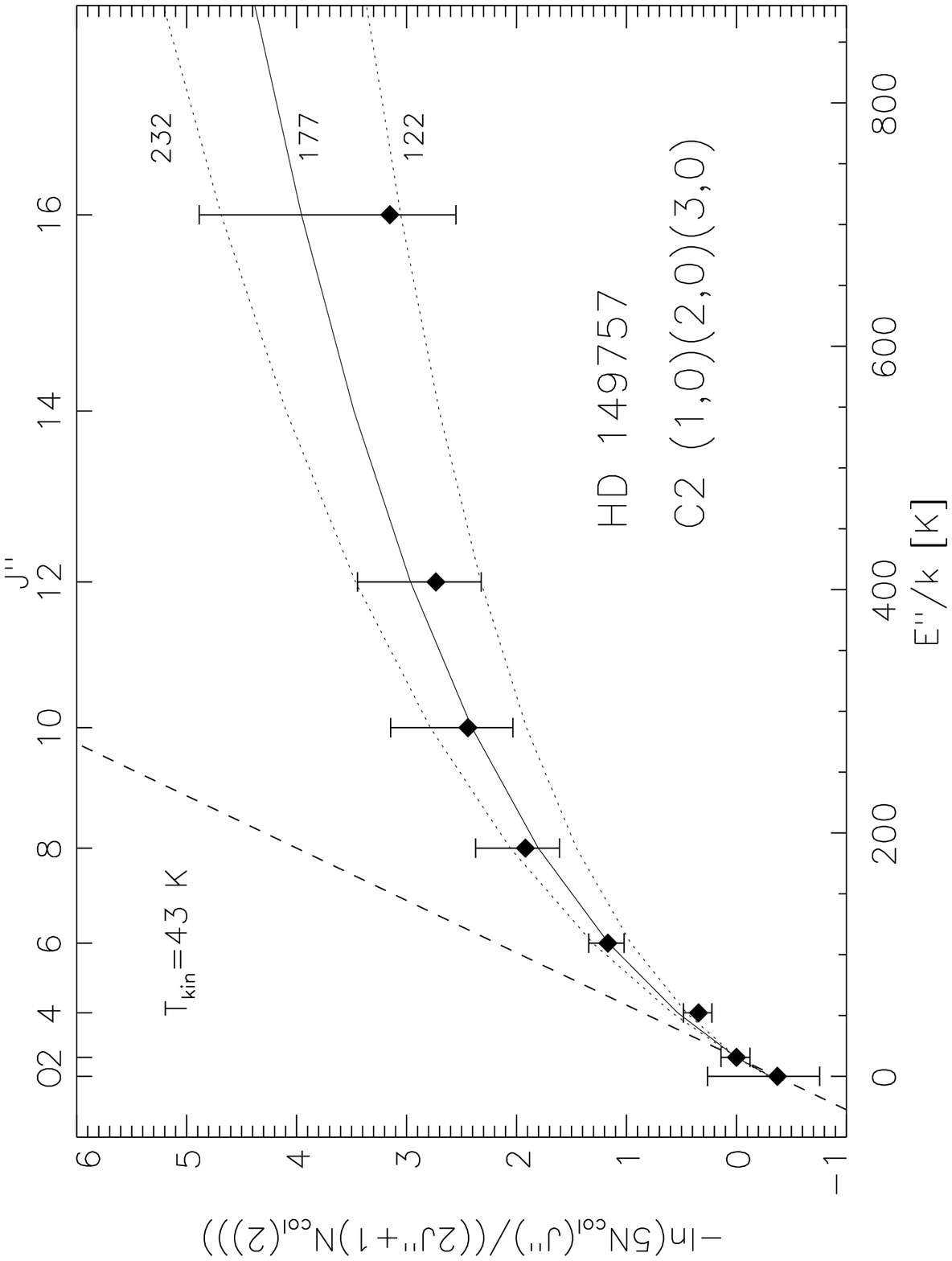}
\includegraphics[width=0.31\textwidth,angle=180]{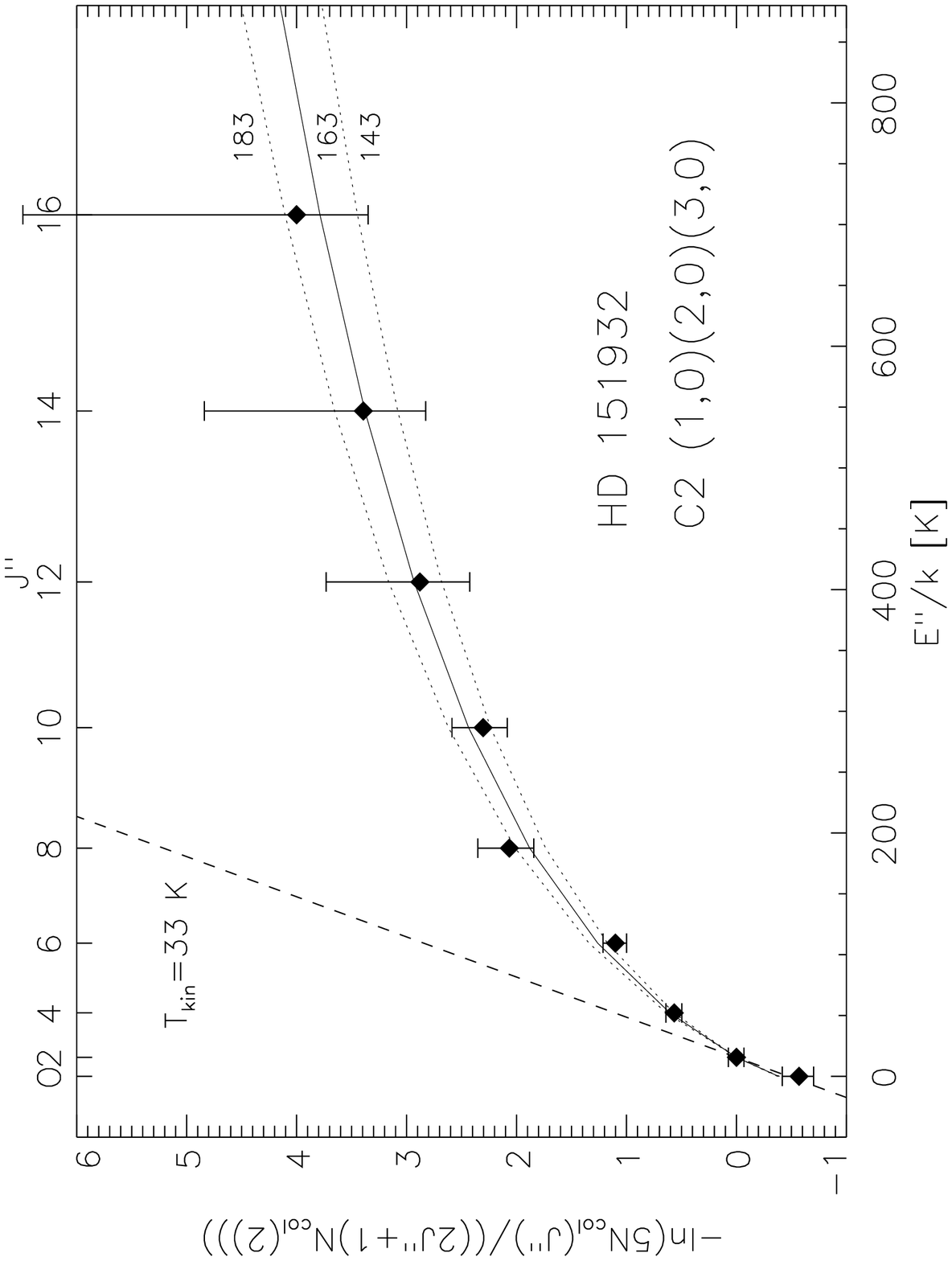}
\includegraphics[width=0.31\textwidth,angle=180]{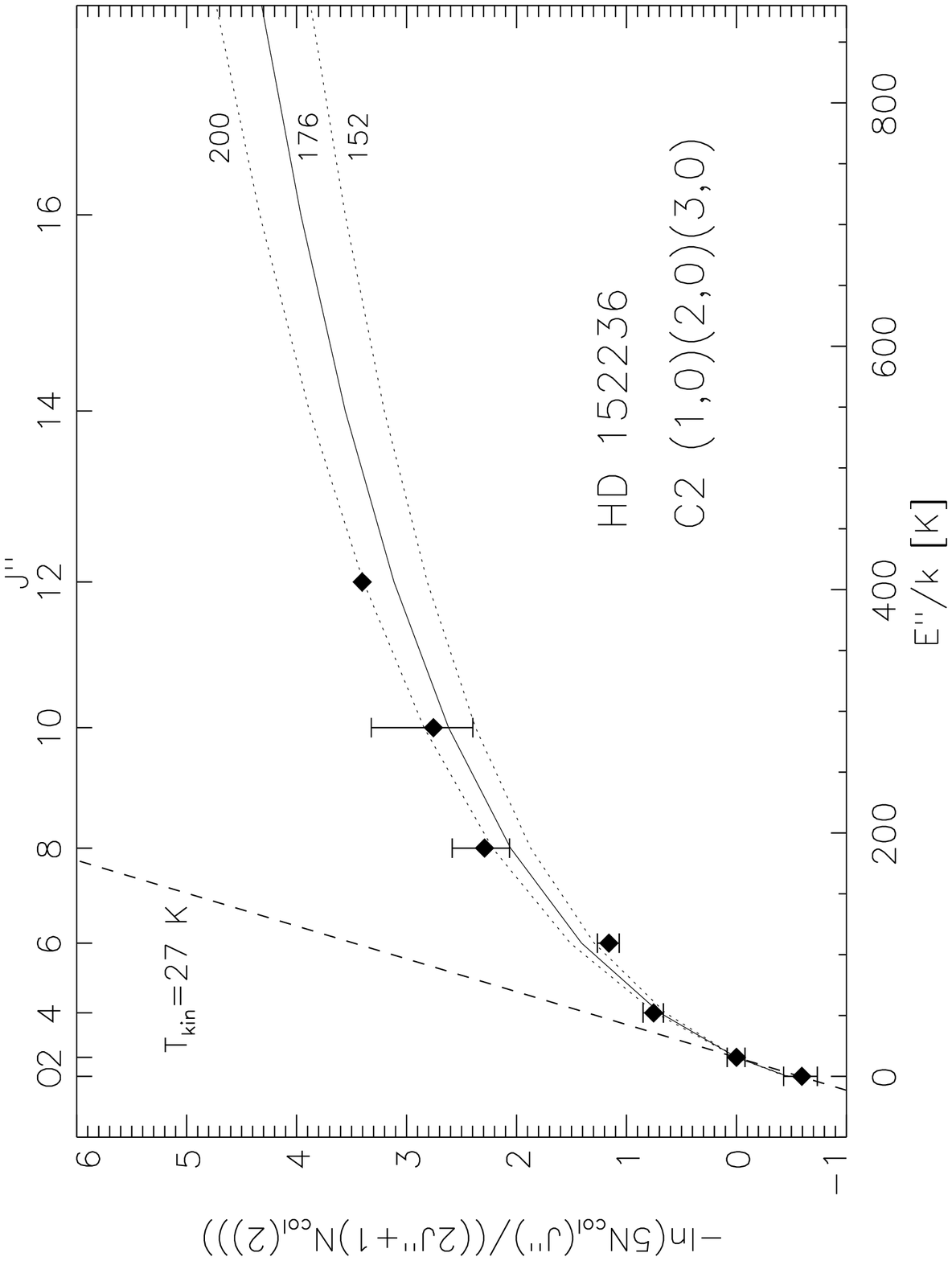}}}
\centerline{
\hbox{
\includegraphics[width=0.31\textwidth,angle=180]{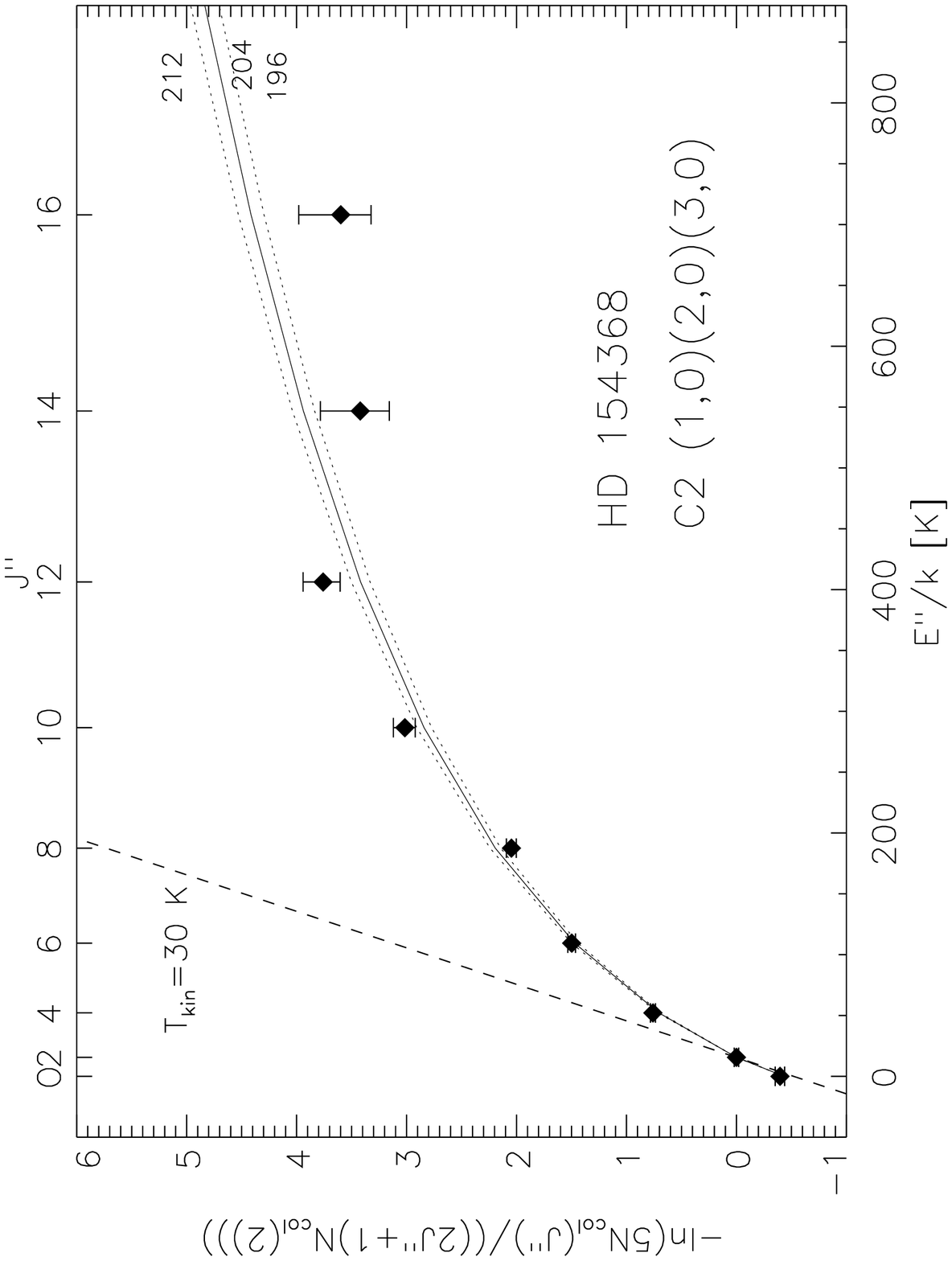}
\includegraphics[width=0.31\textwidth,angle=180]{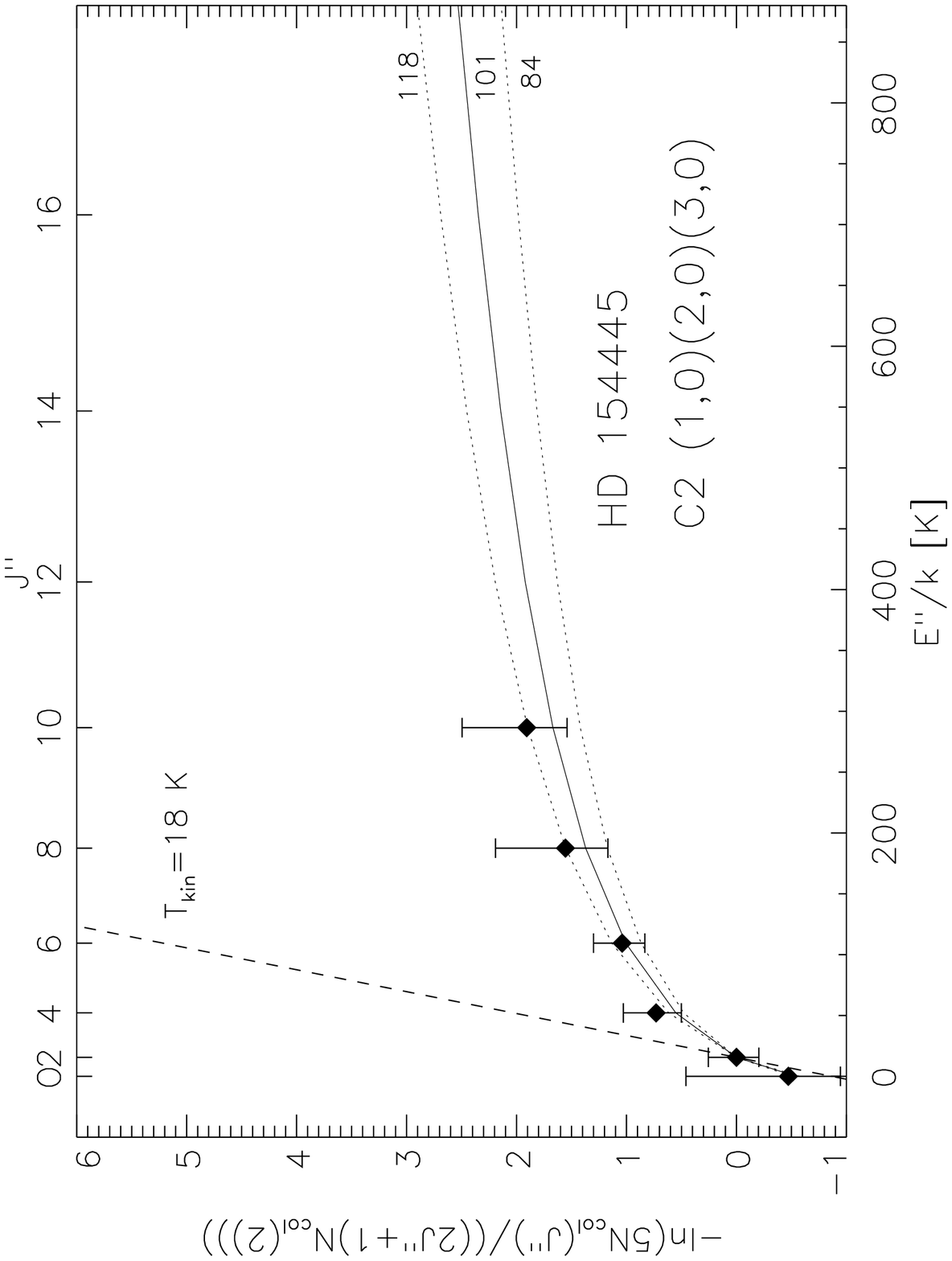}
\includegraphics[width=0.31\textwidth,angle=180]{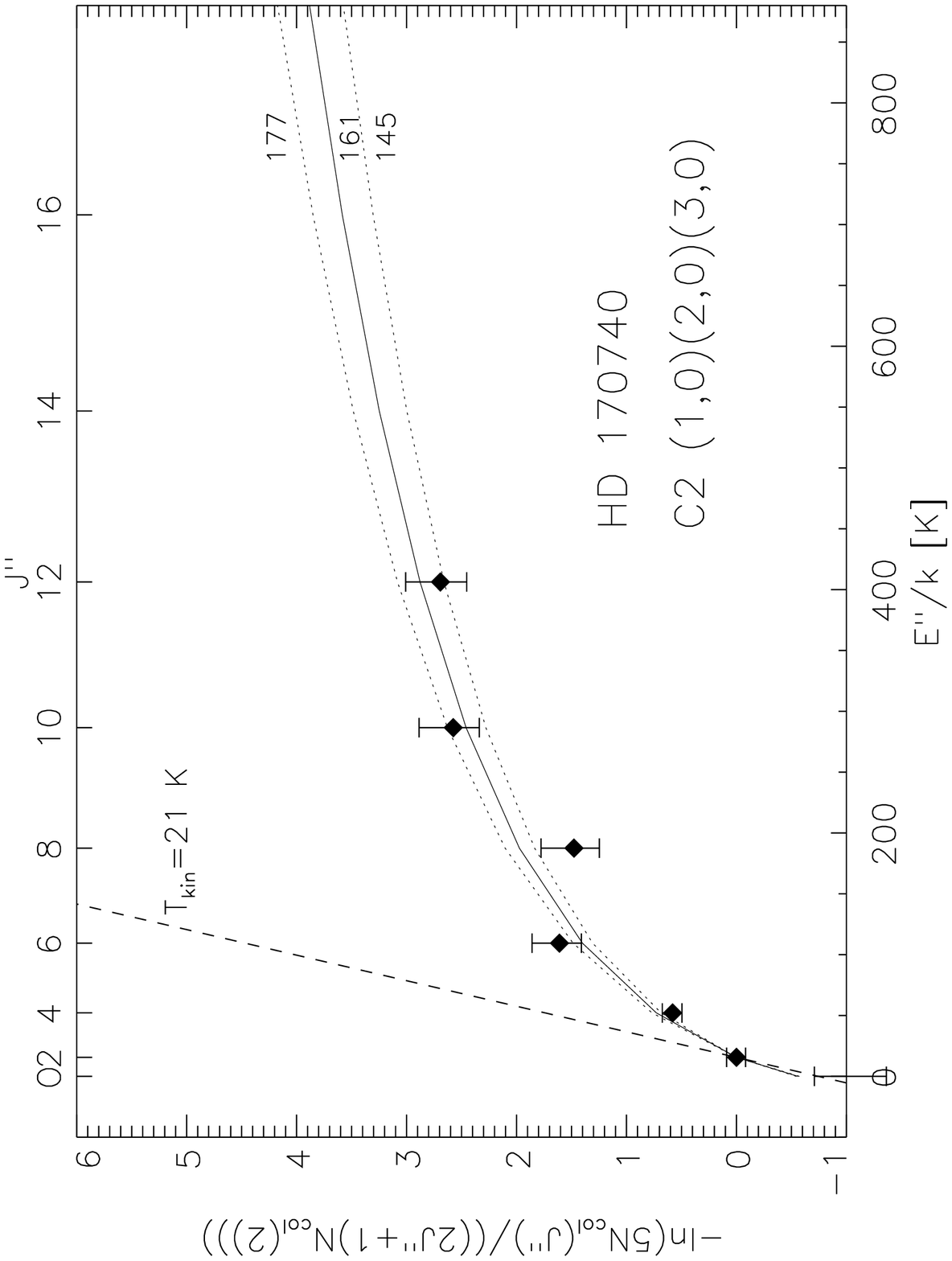}}}
\caption{Relative $C_2$ rotational population diagrams toward six program stars, as a function of the excitation energy (and rotational quantum number $J''$). Following van Dishoeck \& Black (1982) we computed grid of models for the interpretation of the rotational diagrams and calculation gas kinetic temperature $T_{kin}$ and collisional partners density $n_c = n_H + n_{H_2}$. Hence the solid lines represent fit to the theoretical model, based on the analysis of van Dishoeck \& Black (1982) where the labels describe appropriate values of $n_c$. The straight dashed line shows the best-fitting $T_{kin}$.}
\label{fig01}
\end{figure*}

We also derived a set of rotational temperatures: $T_{02}$, $T_{04}$, $T_{06}$ corresponding to the mean excitation temperatures derived from a linear fit to logarithm of column densities of the first: two, three and four levels, respectively, starting from \mbox{$J''= 0$} (Table 5).

\section{DIBs}
\subsection{Observational data}

\begin{figure*}
\centerline{
\hbox{
\includegraphics[width=0.27\textwidth]{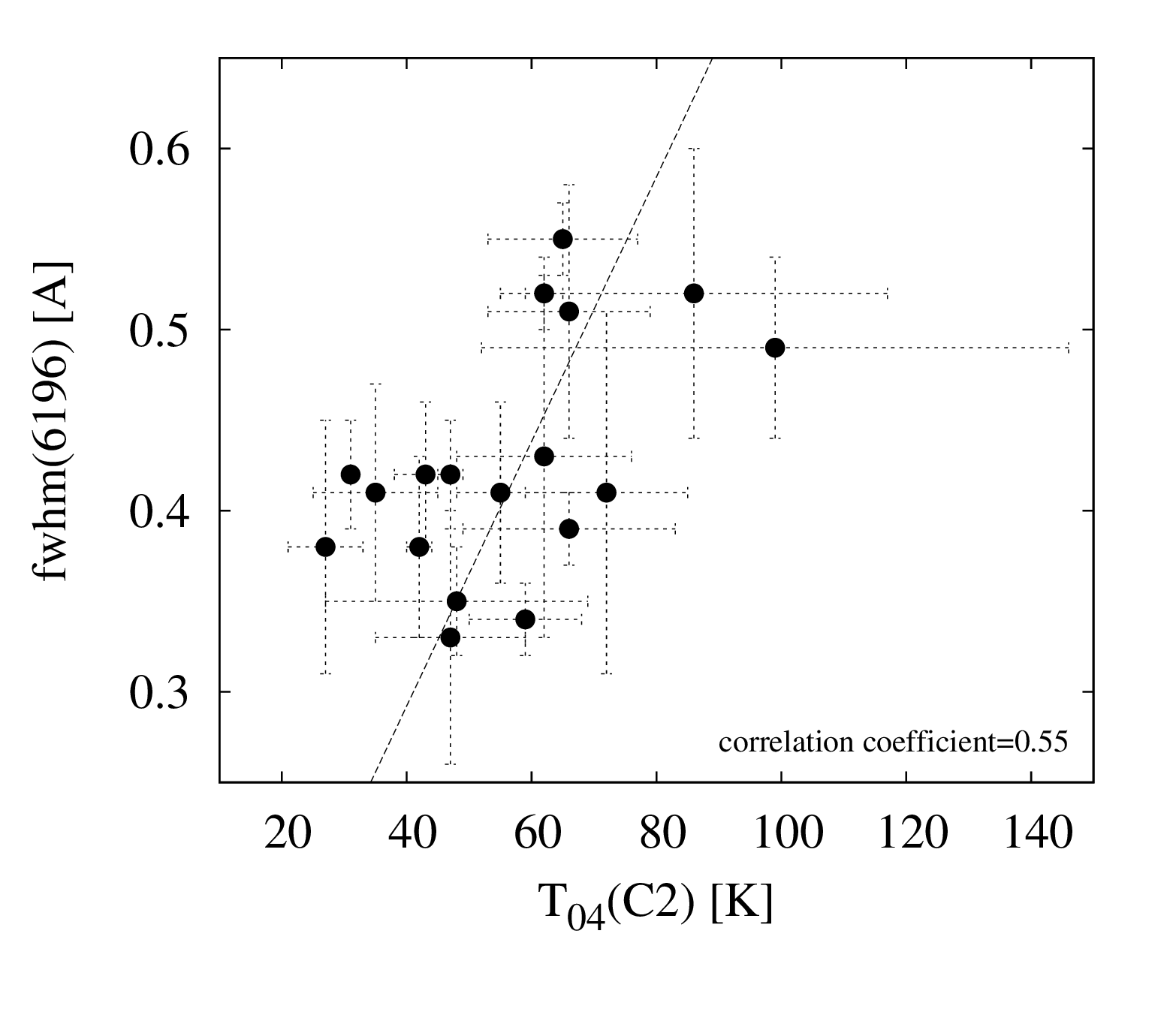}
\includegraphics[width=0.27\textwidth]{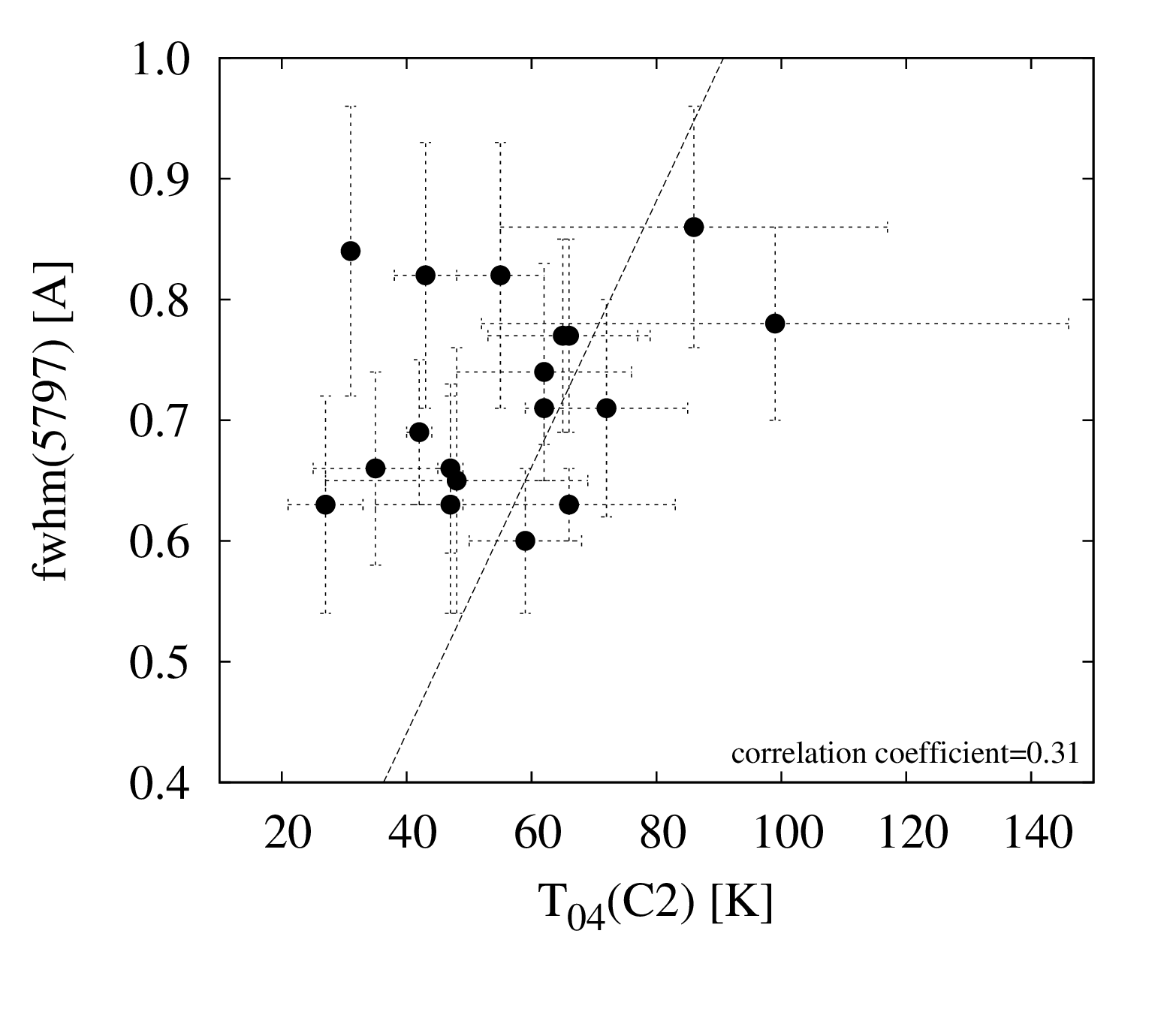}
\includegraphics[width=0.27\textwidth]{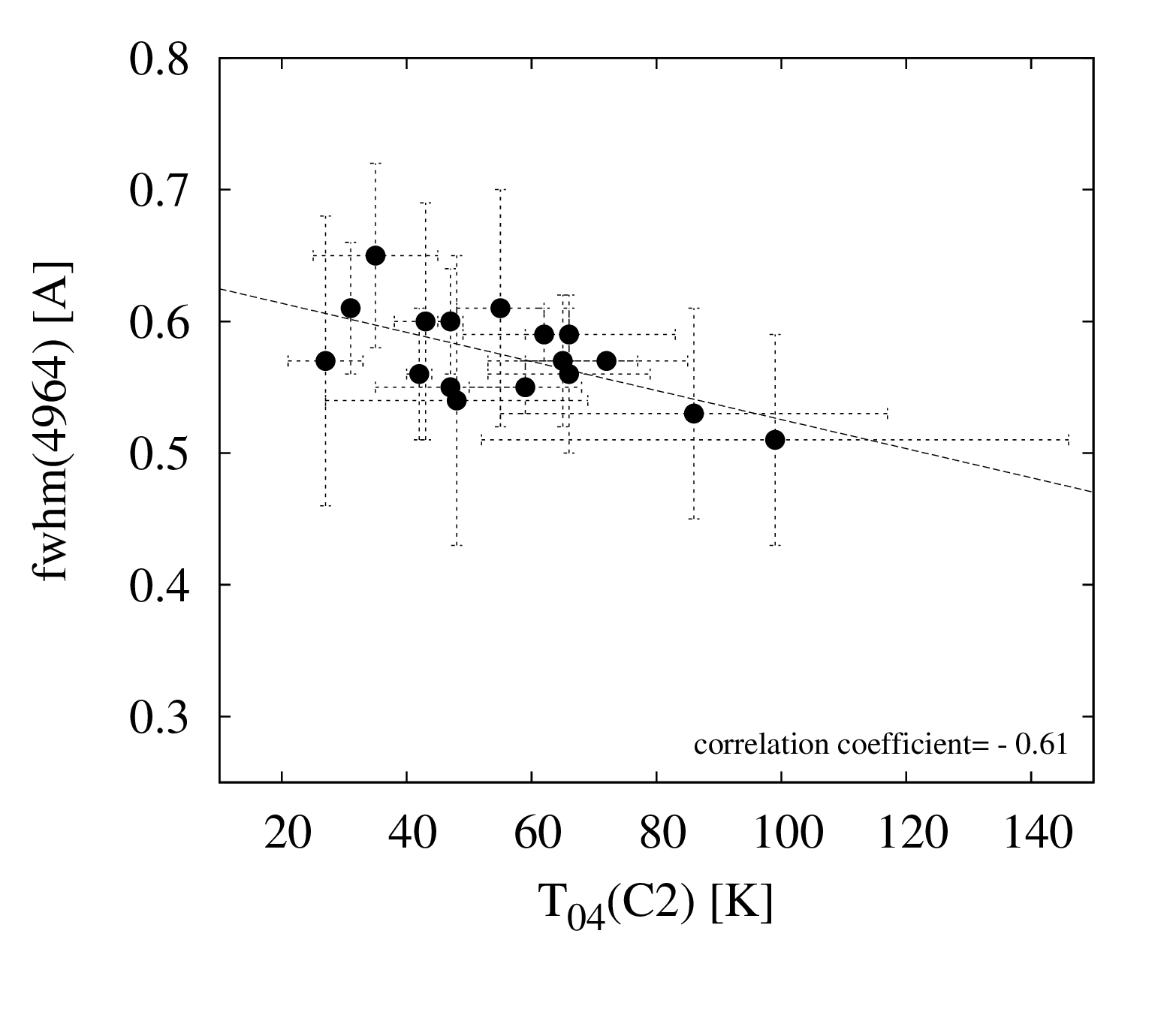}
\includegraphics[width=0.27\textwidth]{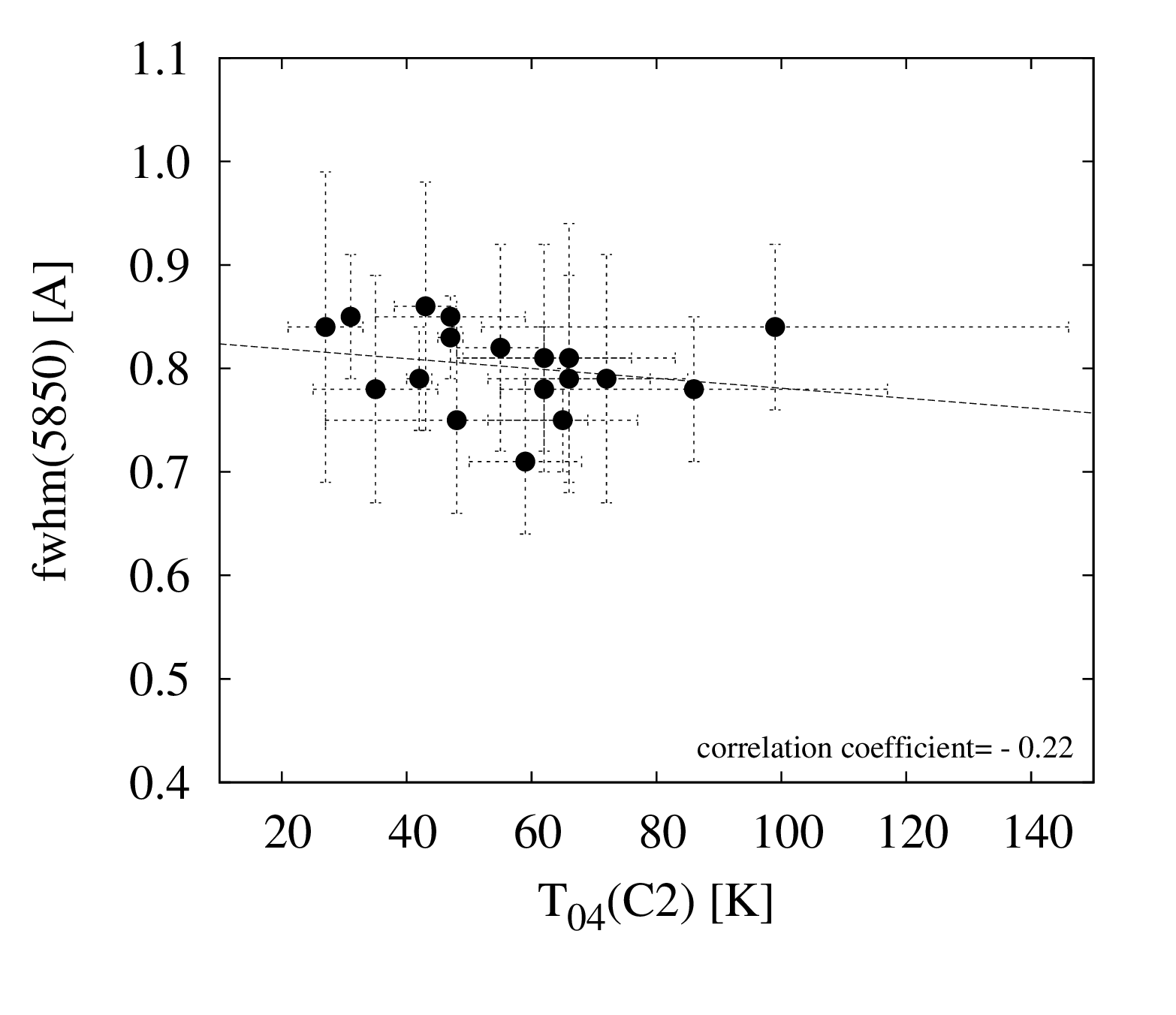}
}}
\vspace{0.1cm}
\centerline{
\hbox{
\includegraphics[width=0.27\textwidth]{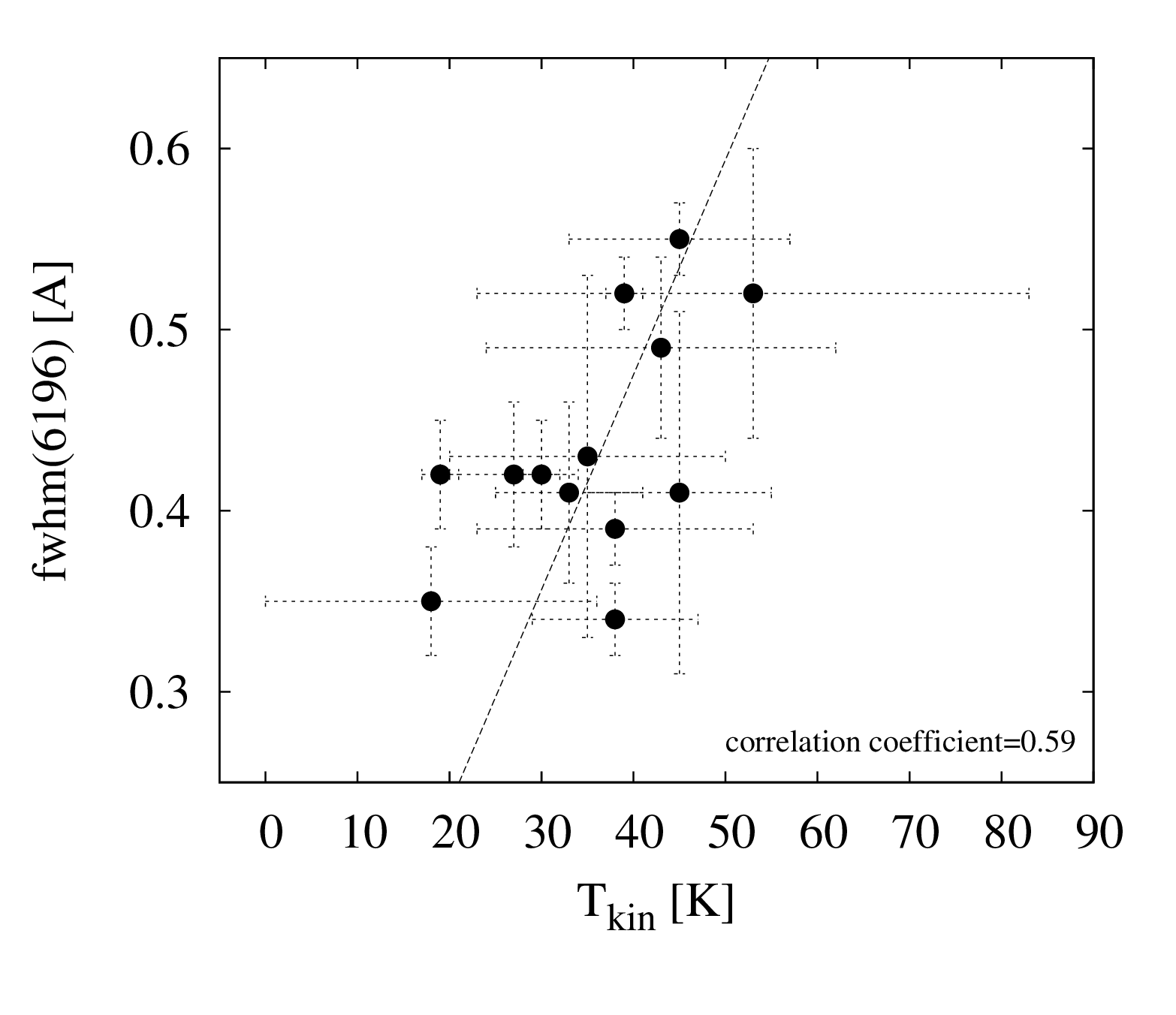}
\includegraphics[width=0.27\textwidth]{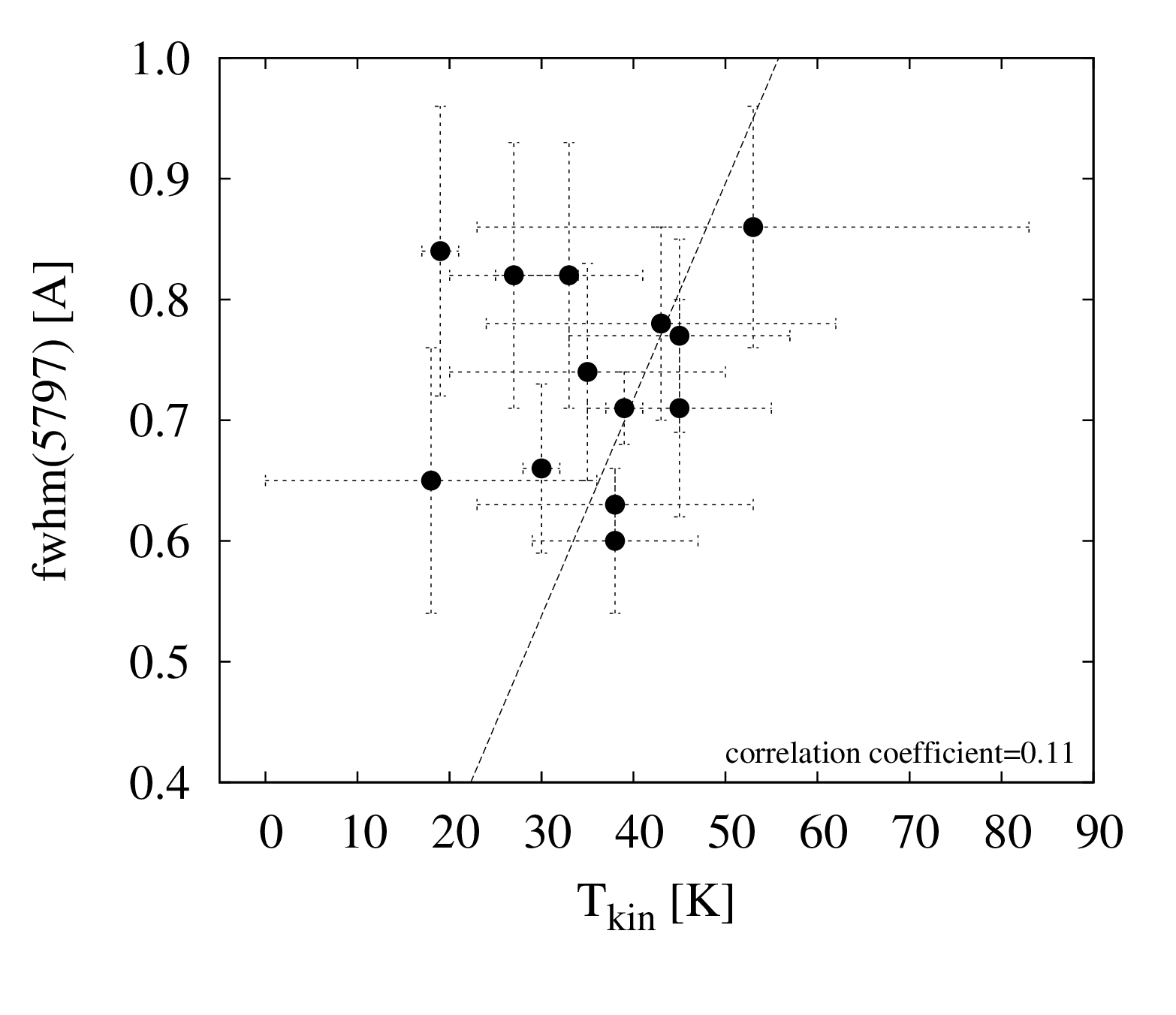}
\includegraphics[width=0.27\textwidth]{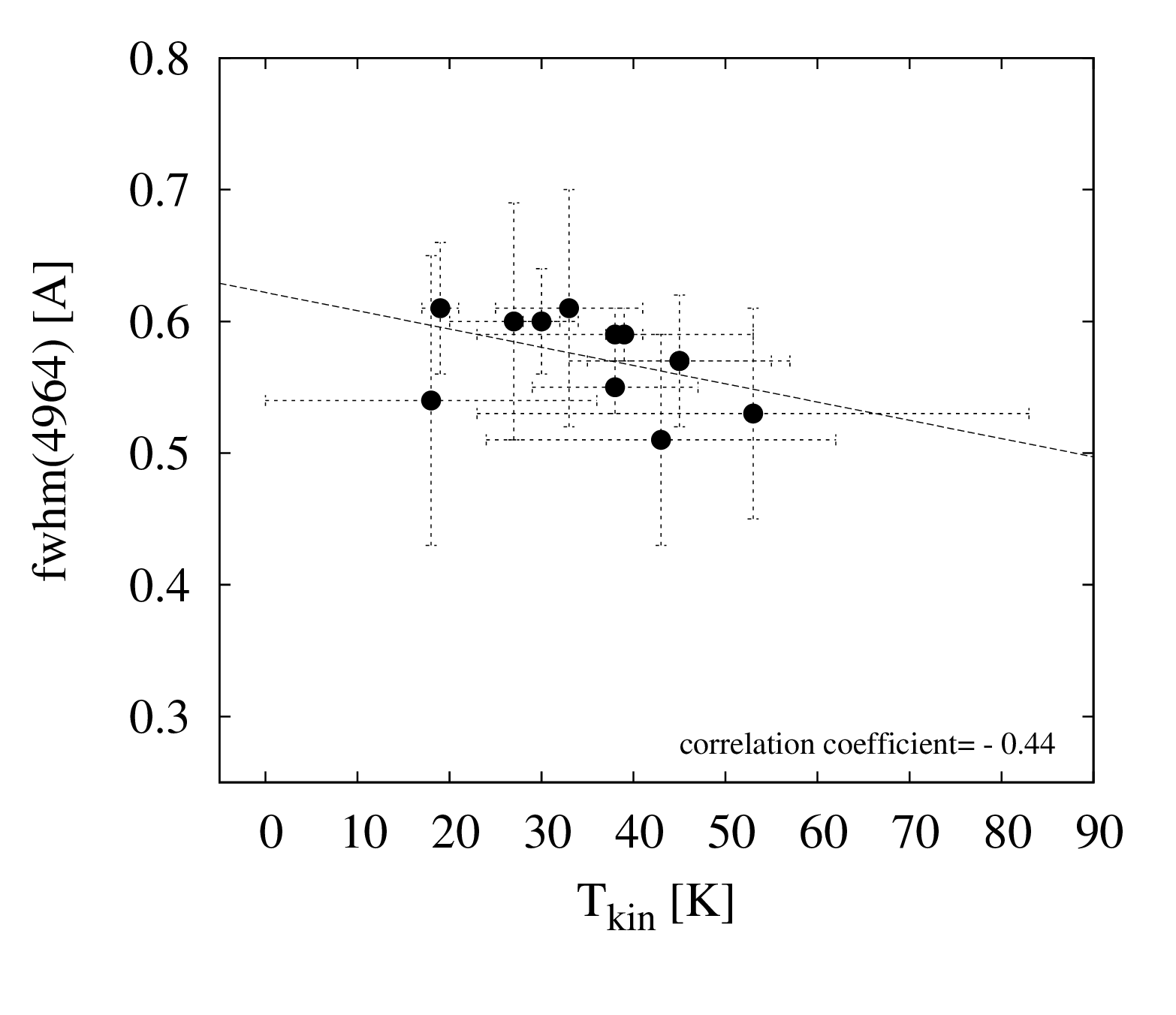}
\includegraphics[width=0.27\textwidth]{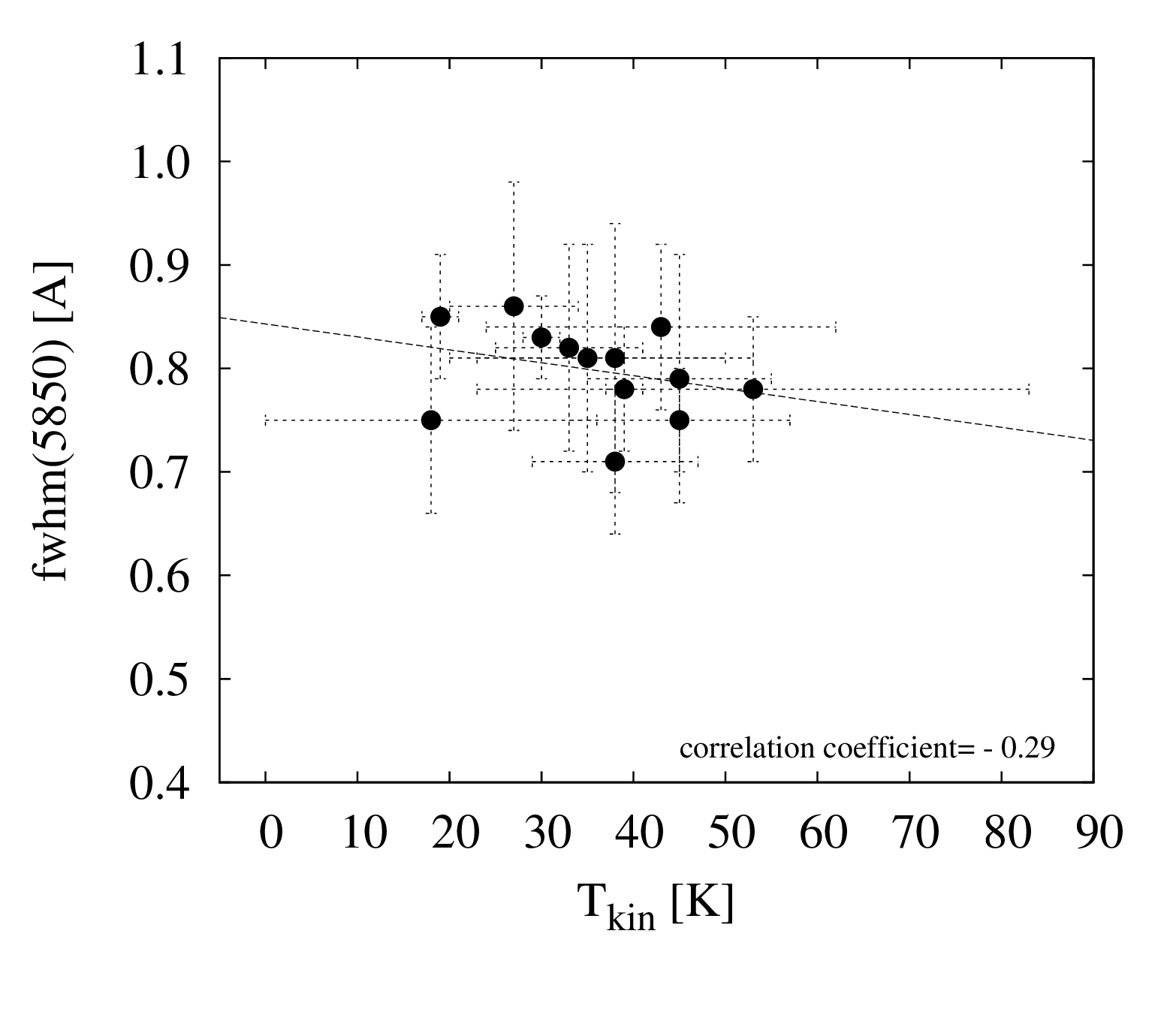}
}}
\caption{The comparison of the FWHM of DIBs' 6196\AA, 5797\AA, 4964\AA,
5850\AA\, to $T_{04}(C_2)$ (the upper panel) and $T_{kin}$ (the bottom panel) toward programme objects.
The straight lines show the best fitting for the measurements; the correlation coefficients are written in the right bottom corners of each panel.
The profile width of the DIB 6196\AA\, and 5797\AA\, varie's from object to object,
likely because of the varying excitation temperature $T_{04}$ of $C_2$
(and also gas kinetic temperature $T_{kin}$), is being wider for higher temperatures.
This effect does not exists in the profiles width of the DIB's 4964\AA\, and 5850\AA.}
\label{fig02}
\end{figure*}

This part of the paper is a continuation of previous work \citep{Kazmierczak2009aa}, where we presented the correlation between the width of a strong diffuse interstellar band at 6196\,\AA~and the excitation temperature of $C_2$. The width and shape of the narrow 6196\,\AA~DIB profile apparently depend on the $C_2$ temperature, being broader for higher values, what can mean that DIB carrier is a centrosymmetric molecule and conditions of excitation of $C_2$ and the DIB 6196\,\AA~carrier should be similar.
\begin{table*}
\caption{Summary of observational data}
\label{results}
\begin{tabular}{lccccccccc}
\hline
object &  FWHM [\AA] & FWHM [\AA] & FWHM [\AA] & FWHM [\AA] &  $T_{02}(C_2)$ & $T_{04}(C_2)$ & $T_{kin}$ & source & source\\
       & (6196)      & (5797)     & (4964)     & (5850)     &        [K]     &      [K]      &     [K]   &$C_2$   & DIB's \\
\hline
HD\,23180  & $0.41\pm0.06$ & $0.66\pm0.08$ & $0.65\pm0.07$ & $0.78\pm0.11$ & $20\pm7$ & $35\pm10$ &          & 1-BOES & BOES\\
HD\,24534  & $0.41\pm0.10$ & $0.71\pm0.09$ & $0.57\pm0.07$ & $0.79\pm0.12$ &          & $72\pm13$ & $45\pm10$& 2      & MAESTRO\\
HD\,110432 & $0.38\pm0.07$ & $0.63\pm0.09$ & $0.57\pm0.11$ & $0.84\pm0.15$ & $13\pm4$ & $27\pm 6$ &          & 3      & UVES\\
HD\,115842 & $0.52\pm0.08$ & $0.86\pm0.10$ & $0.53\pm0.08$ & $0.78\pm0.07$ & $22\pm9$ & $86\pm31$ & $53\pm30$& 4-UVES & UVES\\
HD\,147888 & $0.51\pm0.07$ & $0.77\pm0.08$ & $0.56\pm0.06$ & $0.79\pm0.10$ & $38\pm20$& $66\pm13$ &          & 2      & FEROS\\
HD\,147889 & $0.52\pm0.02$ & $0.71\pm0.03$ & $0.59\pm0.02$ & $0.78\pm0.06$ & $49\pm7$ & $62\pm 3$ & $39\pm2$ & 5-UVES & HARPS\\
HD\,148184 & $0.55\pm 0.02$& $0.77\pm0.08$ & $0.57\pm0.05$ & $0.75\pm0.05$ & $82\pm82$& $65\pm12$ & $45\pm12$& 5-UVES & HARPS\\
HD\,149757 & $0.49\pm0.05$ & $0.78\pm0.08$ & $0.51\pm0.08$ & $0.84\pm0.08$ & $42\pm42$& $99\pm47$ & $43\pm19$& 4-UVES & UVES\\
HD\,151932 & $0.41\pm0.05$ & $0.82\pm0.11$ & $0.61\pm0.09$ & $0.82\pm0.10$ & $28\pm7$ & $55\pm7$  & $33\pm8$ & 4-UVES & UVES\\ 
HD\,152236 & $0.42\pm0.04$ & $0.82\pm0.11$ & $0.60\pm0.09$ & $0.86\pm0.12$ & $26\pm8$ & $43\pm5$  & $27\pm7$ & 4-UVES & UVES\\
HD\,154368 & $0.42\pm0.03$ & $0.66\pm0.07$ & $0.60\pm0.04$ & $0.83\pm0.04$ & $39\pm5$ & $47\pm2$  & $30\pm2$ & 4-UVES & UVES\\
HD\,154445 & $0.35\pm0.03$ & $0.65\pm0.11$ & $0.54\pm0.11$ & $0.75\pm0.09$ & $33\pm33$& $48\pm21$ & $18\pm18$& 4-UVES & UVES\\
HD\,163800 & $0.39\pm0.02$ & $0.63\pm0.03$ & $0.59\pm0.02$ & $0.81\pm0.13$ & $24\pm 8$& $66\pm17$ & $38\pm15$& 5-UVES & HARPS\\
HD\,169454 & $0.42\pm0.03$ & $0.84\pm0.12$ & $0.61\pm0.05$ & $0.85\pm0.06$ & $23\pm 2$& $31\pm1 $ & $19\pm2$ & 5-UVES & UVES\\
HD\,179406 & $0.34\pm0.02$ & $0.60\pm0.06$ & $0.55\pm0.02$ & $0.71\pm0.07$ & $38\pm12$& $59\pm9 $ & $38\pm9$ & 5-UVES & HARPS\\
HD\,204827 & $0.38\pm0.05$ & $0.69\pm0.06$ & $0.56\pm0.05$ & $0.79\pm0.05$ & $34\pm34$& $42\pm 2$ &          & 6      & BOES\\
HD\,207538 & $0.43\pm0.10$ & $0.74\pm0.09$ &               & $0.81\pm0.11$ &          & $62\pm14$ & $35\pm15$& 7      & MAESTRO\\
HD\,210839 & $0.33\pm0.07$ & $0.63\pm0.09$ & $0.55\pm0.07$ & $0.85\pm0.12$ & $16\pm6$ & $47\pm12$ &          & 2      & MAESTRO\\
\hline
\multicolumn{10}{l}{FWHM's of DIB's were measured in this paper;}\\
\multicolumn{10}{l}{1 - Ka{\'z}mierczak et al. (2009), 2 - Sonnentrucker et al (2007), 3 - Sonnentrucker et al. (2007) using the EW's of van Dishoeck}\\
\multicolumn{10}{l}{\& Black (1989), 4 - this paper, 5 - Ka{\'z}mierczak et al. (2010), 6 - Ka{\'z}mierczak et al. (2009) based on EW's from {\'A}d{\'a}mkovics}\\
\multicolumn{10}{l}{et al. (2003), 7 - Sonnentrucker et al. (2007) using the EW's of Galazutdinov et al. (2006)}
\end{tabular}
\end{table*}

Since that time the sample of interstellar clouds where a detailed analysis of excitation of $C_2$ was made, was increased and we analysed three more DIBs (4964\,\AA, 5850\,\AA~and 5797\,\AA). 

Spectra which were used to this purpose are the same as in \citep{Kazmierczak2009aa} plus new data described in Section 2.
In column 9 and 10 in Table 6 information about data source\footnote{
BOES - Bohyunsan Optical Astronomical Observatory, South Korea 1.8m;
MAESTRO - International Centre for Astronomical and Medico-Ecological Research, Terskol, Russia, 2m;
FEROS - Fiber-fed Extended Range Optical Spectrograph, ESO La Silla, Chile 2.2m;
UVES - UV-Visual Echelle Spectrograph, ESO Paranal, Chile, 8m;
HARPS - High Accuracy Radial velocity Planet Searcher, ESO La Silla, Chile 3.6m.} is given (for more details - see \citet{Kazmierczak2009aa}).
 
Table 6 summarised all results for 18 interstellar clouds where $C_2$ was observed and where the full widths at half maximum FWHM of four DIBs (4964\,\AA, 5850\,\AA, 5797\,\AA~and 6196\,\AA) were quite easy to measure. 
The rotational temperature of $C_2$ varies from object to object; gas kinetic temperature also shows that tendency but the density of collision partners in a molecular cloud $n_c$ is more or less $10^2$ ${cm}^{-3}$ for all of the objects. 

\section{Results and discussion}

We have analysed only four DIBs (6196, 5797, 4964 and 5850), because the other ones are usually very broad features, what makes them difficult for measurements. Quite strong and narrow diffuse interstellar bands were chosen to show that some of them are well correlated with excitation temperature of $C_2$ and the others are not. These four features are enough to evidently show that tiny effect. 
It is also interesting that DIBs 6196\,\AA~and 5797\,\AA~for which that effect exist, were classified as 'NO-C2-DIB' by Thourburn et al. (2003); the same authors said that DIB at 5850\,\AA~is 'C2-DIB', but we showed there is no correlation with temperature of $C_2$.

Currently we know 33 interstellar clouds where the detailed analysis of $C_2$ was made (24 - Sonnentrucker et al 2007, 2 - Ka{\'z}mierczak et al 2010, 7 - this paper). Not all of them were used to analyse DIBs because we do not have all spectra and in some of them $CH$ or $KI$ have more than one Doppler component. In that case weak $C_2$ lines can be measured, but broad DIB features could be contaminated by multi-components from different interstellar clouds in one line of sight.

The main result is shown in Figure 2 (see also Figs 3 and 4). The correlation between the full width at half maximum of the DIBs and the excitation temperature ($T_{04}$) of $C_2$ and gas kinetic temperature is presented. 

For DIB's at 6196\,\AA~it was already shown in \citet{Kazmierczak2009aa}, but now the sample of objects is bigger and for some of them data were a bit recalculated because of errors of the equivalent widths (see Sect. 3 of \citep{Kazmierczak2009mnras}). 
However the correlation is evident; the profiles of the DIB's at 6196\,\AA~depend on the excitation temperature of $C_2$ being broader for its higher values. The same situation is for the DIB at 5797\,\AA~(Figure 3). These results may suggest that carriers of some DIBs could be centrosymmetric molecules. This means that conditions of excitation of $C_2$ and the DIBs 6196\,\AA~and 5797\,\AA~carriers should be similar.

There are also DIBs (4964\,\AA, 5850\,\AA~(Fig 4)) for which that effect does not exist. Analogous figures of 4964\,\AA, 5850\,\AA~to 6196\,\AA~or 5797\,\AA~are definitely different. There is no correlation for 4964 and 5850. The same situation is for excitation temperature of $C_2$ and also for gas kinetic temperature. 

In Figure 2 there is also written the correlation coefficient for each relation. For each source, the correlation coefficient of $T_{04}$ is quite similar to that one of $T_{kin}$. Based on them the best relation is for 6196\,\AA~(the correlation coefficient = 0.55 (for $T_{04}$) and 0.59 (for $T_{kin}$). For 5797\,\AA~it is 0.31 and 0.11. These factors are very low and relations are very tiny, but it looks totally different from the relations for 4964\,\AA~and 5850\,\AA. Correlations for these two DIBs are negative (-0.61 and -0.44 for 4964\,\AA; -0.22 and -0.29 for 5850\,\AA). 
{\bf Conclusions based on the correlation coefficients are not so confident, but still we suggest that these DIBs can be divided into two classes. More sightlines would be needed to justify any definite conclusion.}

That effect is very tiny. Let's emphasize that measurements of weak $C_2$ features are very difficult and also measurements of diffuse interstellar bands are not so easy because these features are very broad. That causes big errors for all of the results. Apart from that some correlations between excitation temperature derived from $C_2$ and width of DIBs exist.
Seemingly the carriers of 6196 and 5797 DIBs are centrosymmetric while those of 4964 and 5850 DIBs are polar molecules.

\begin{figure*}
\centerline{
\hbox{
\includegraphics[width=0.3\textwidth]{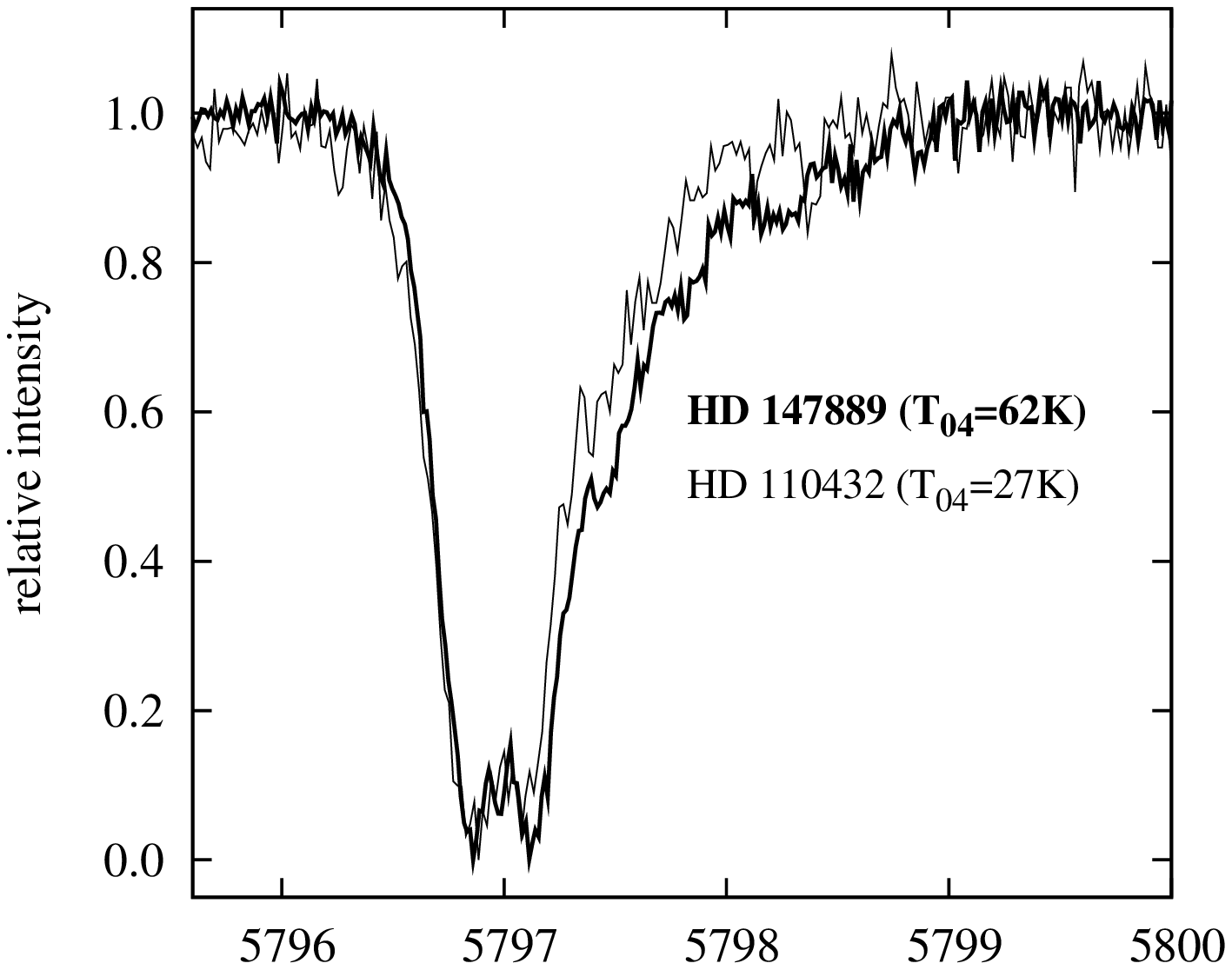}
\includegraphics[width=0.3\textwidth]{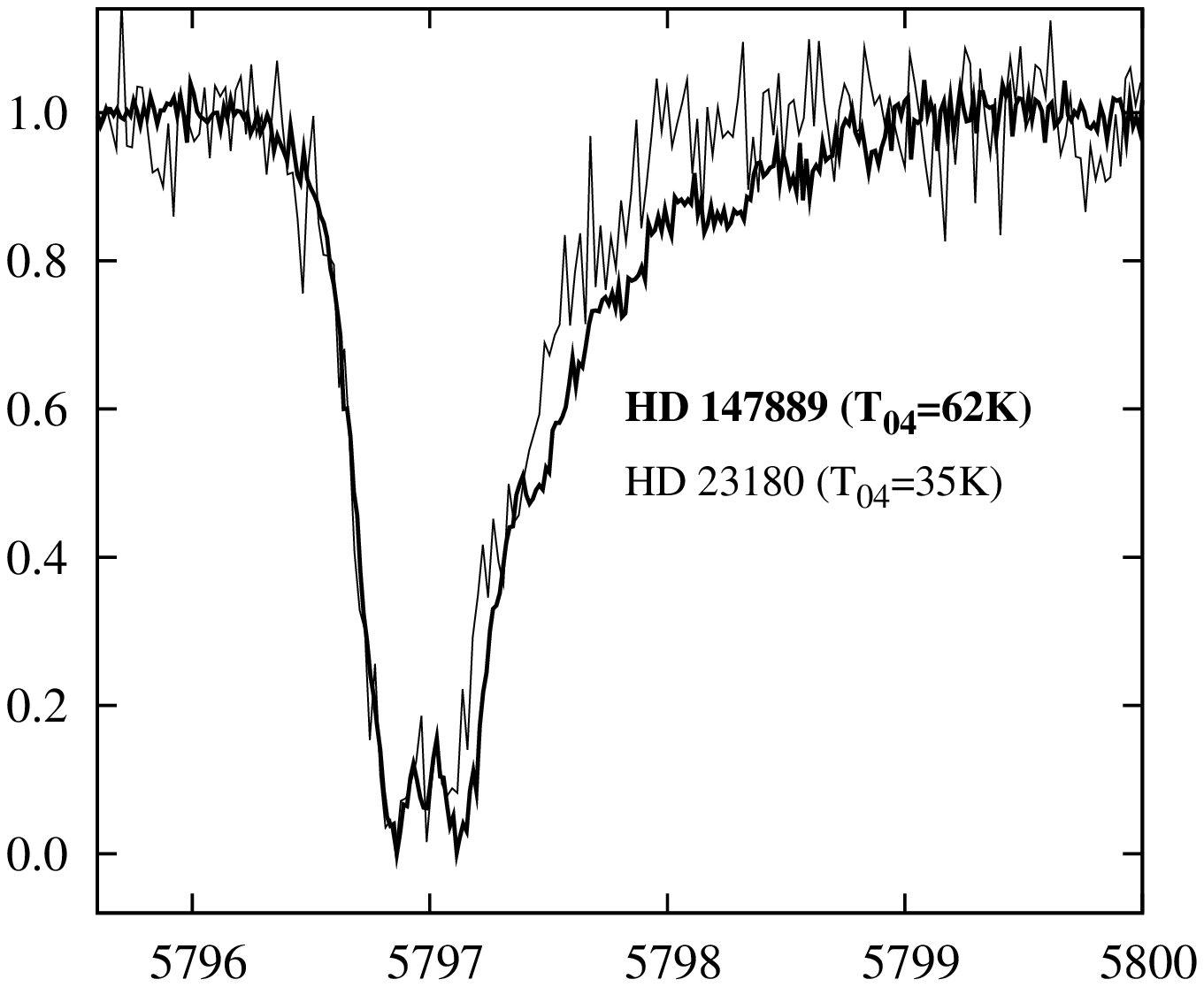}
\includegraphics[width=0.3\textwidth]{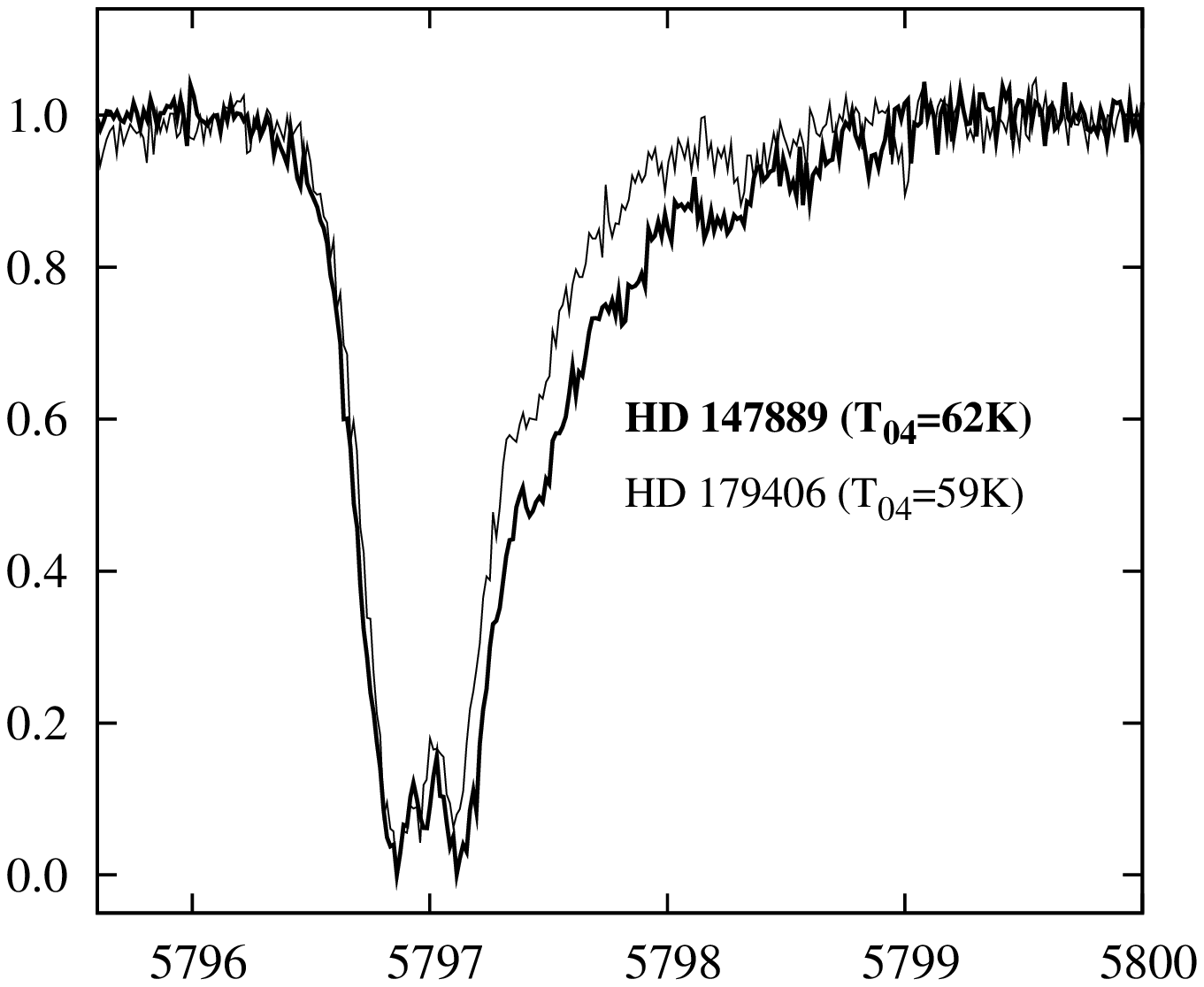}}}
\centerline{
\hbox{
\includegraphics[width=0.3\textwidth]{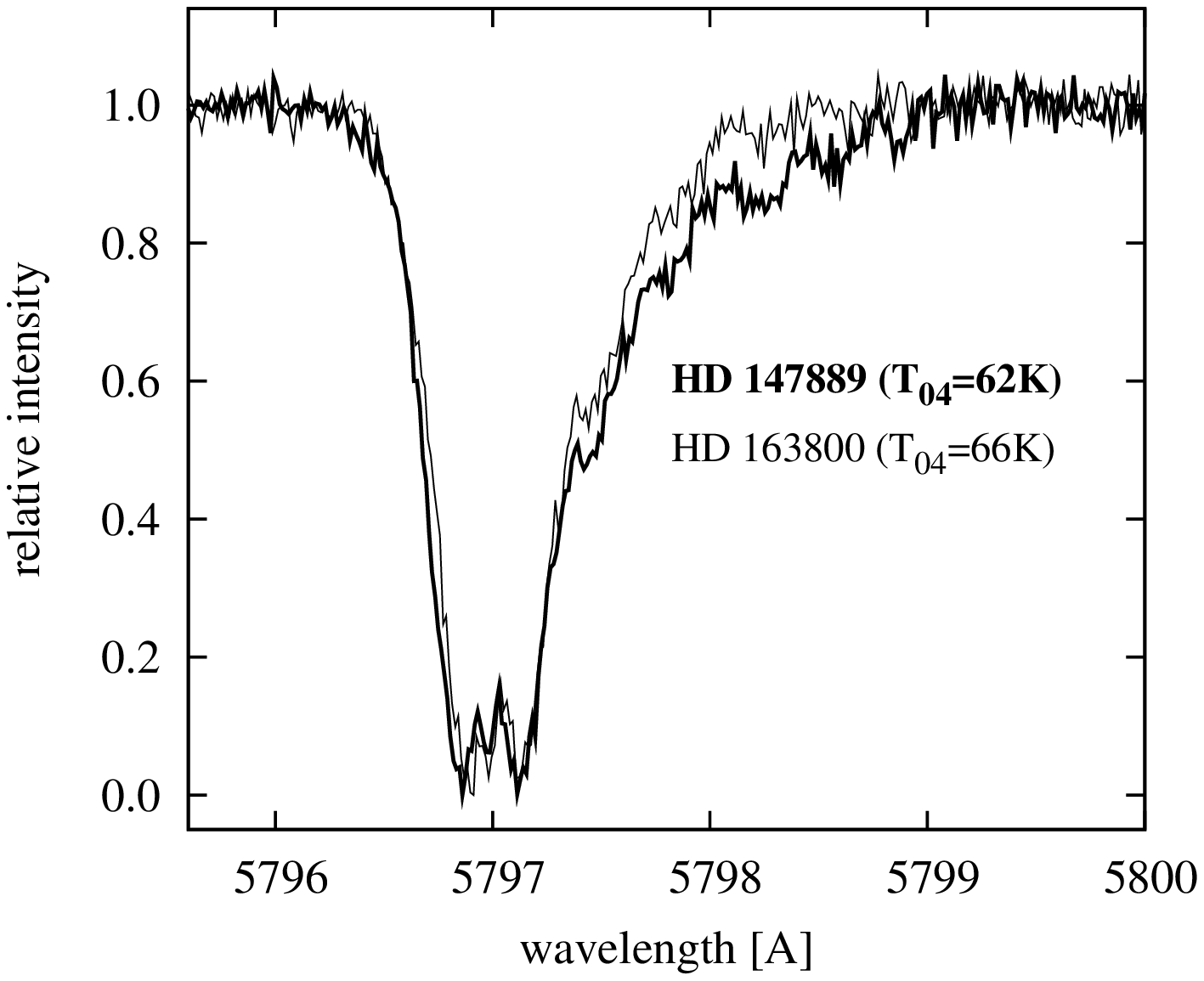}
\includegraphics[width=0.3\textwidth]{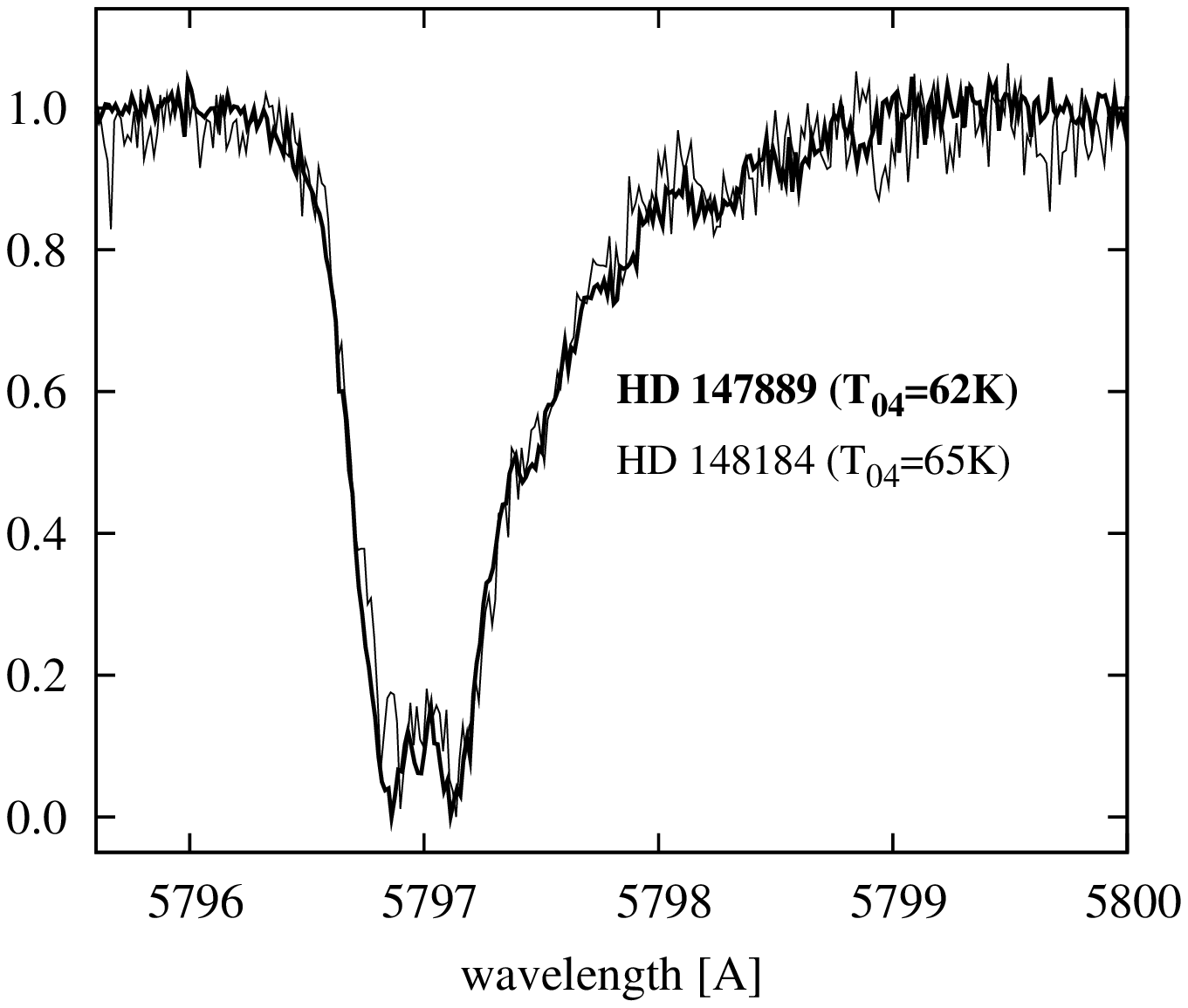}
\includegraphics[width=0.3\textwidth]{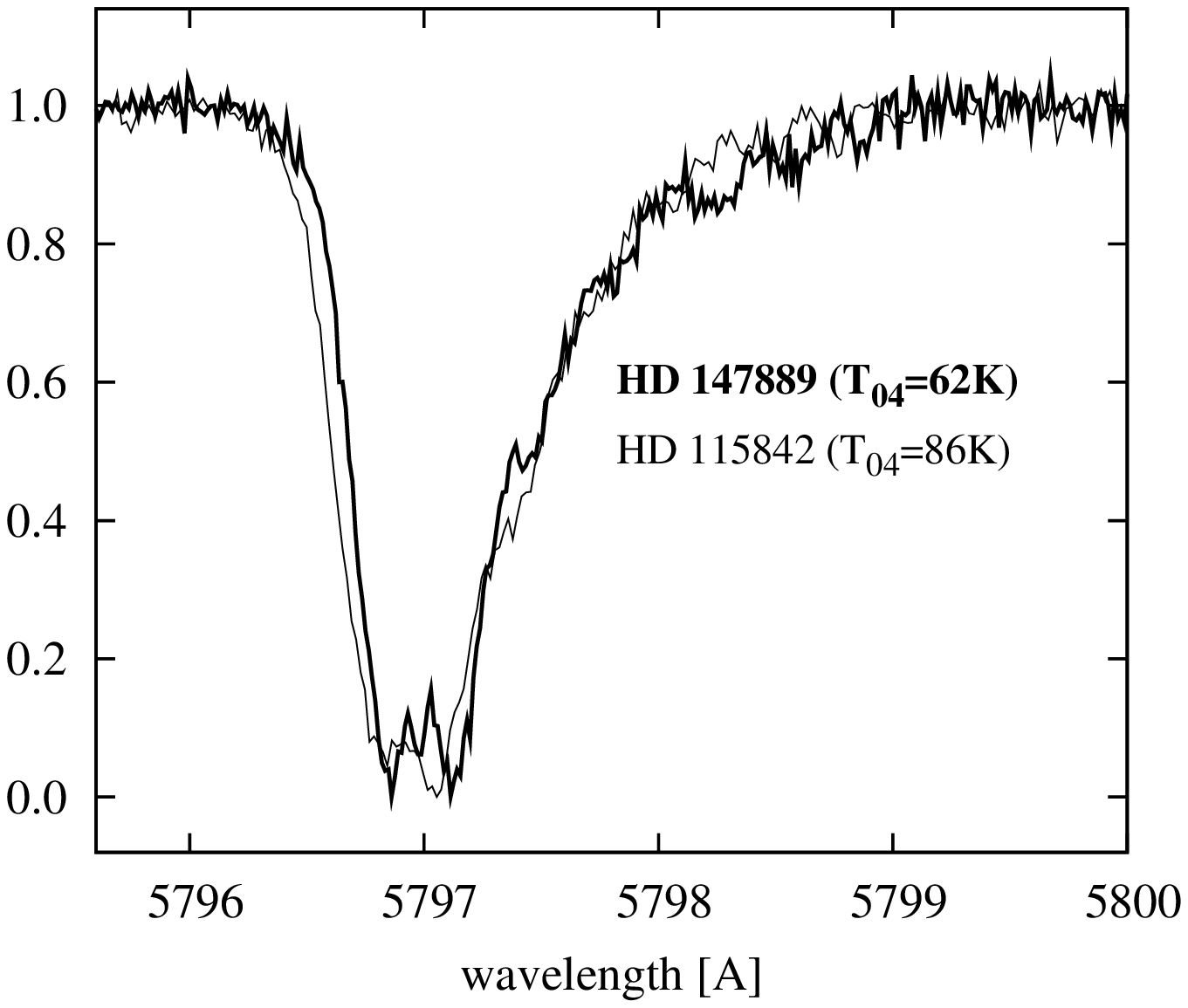}}}
\caption{DIB 5797\,\AA~profiles of the spectra, normalised to their central depths, toward some of the programme stars with different rotational temperatures of $C_2$. DIB profiles depend on the excitation temperature of the dicarbon molecule $T_{04}$.}
\label{fig03}
\end{figure*}

\begin{figure*}
\centerline{
\hbox{
\includegraphics[width=0.3\textwidth]{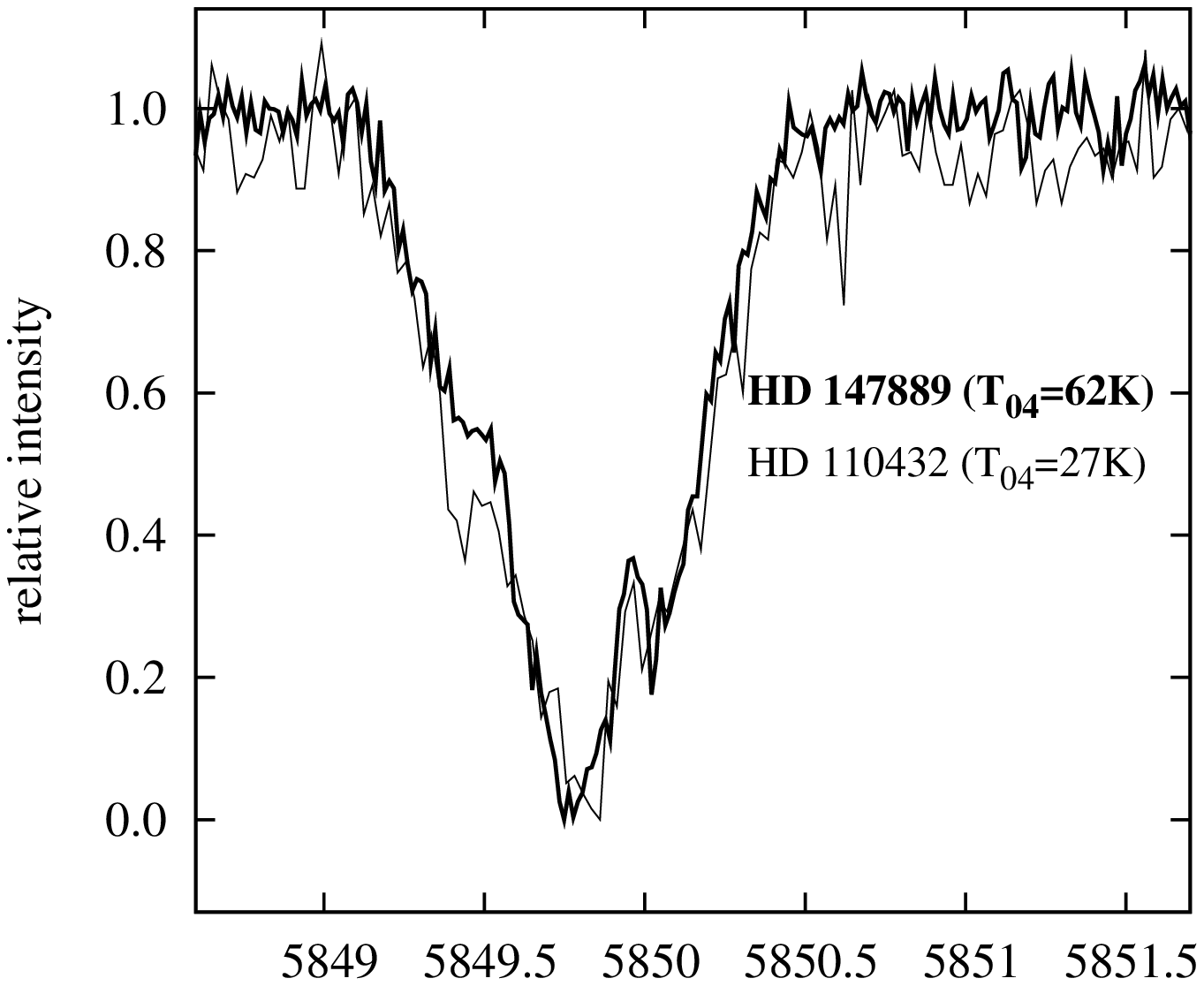}
\includegraphics[width=0.3\textwidth]{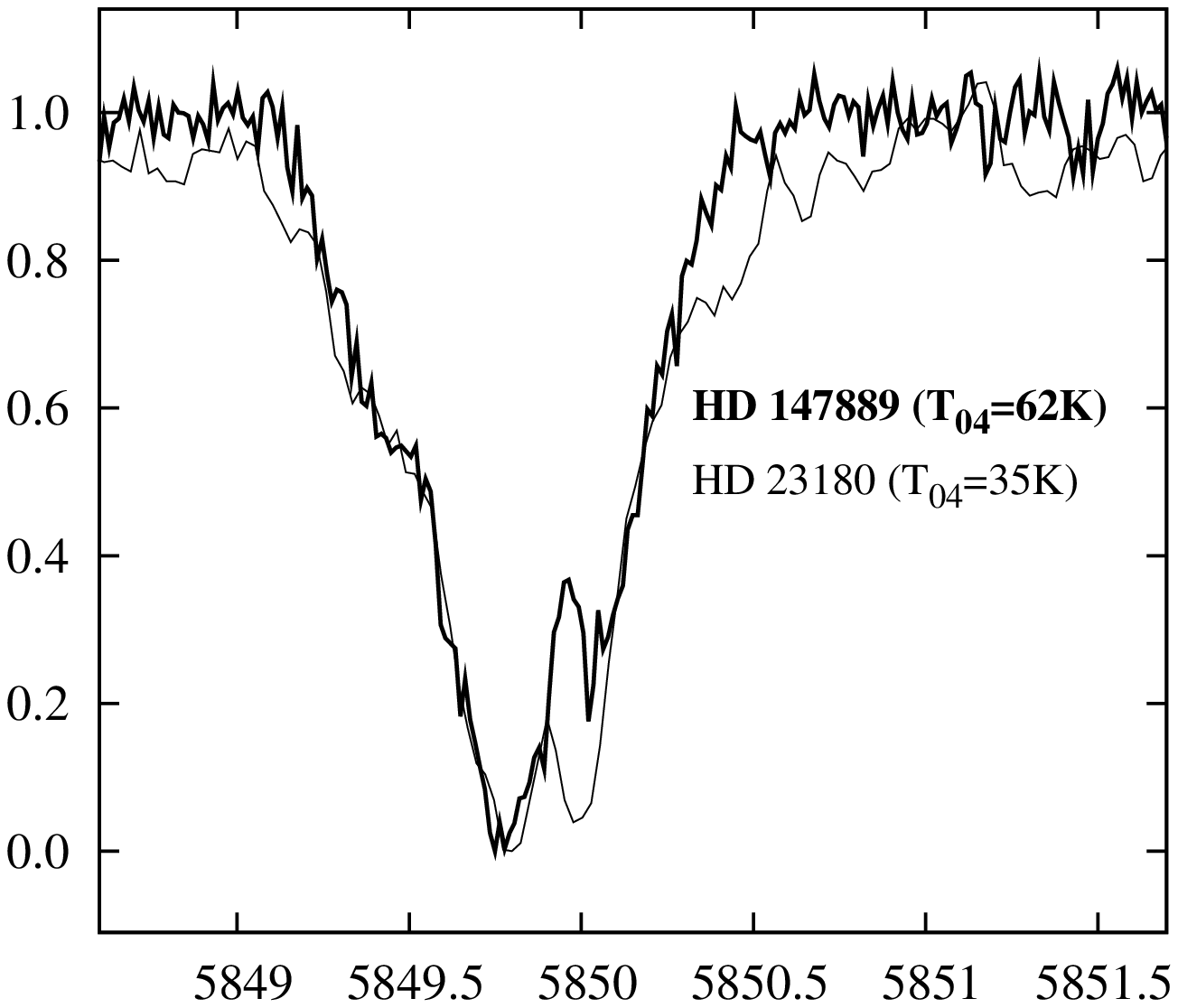}
\includegraphics[width=0.3\textwidth]{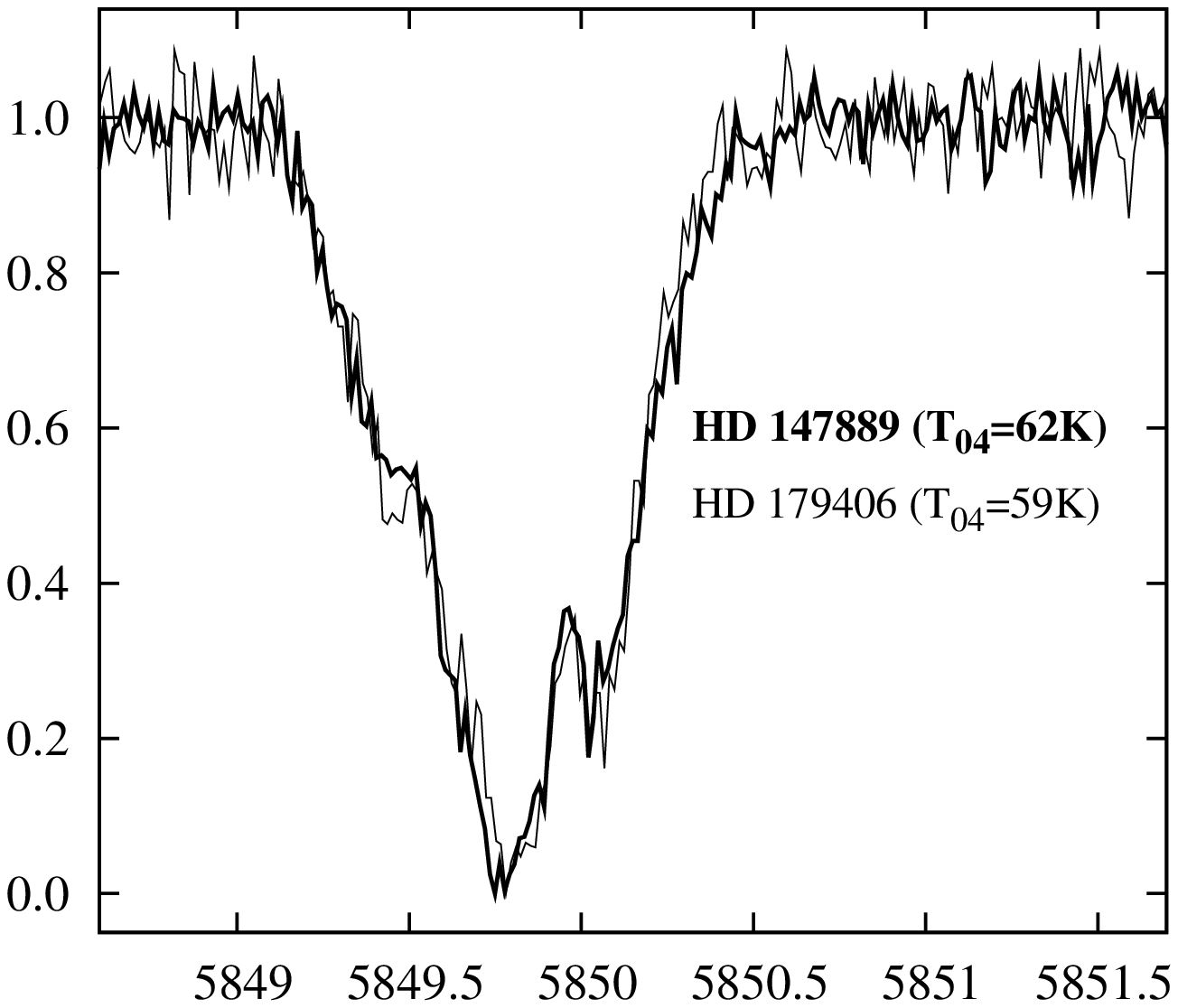}
}}
\centerline{
\hbox{
\includegraphics[width=0.3\textwidth]{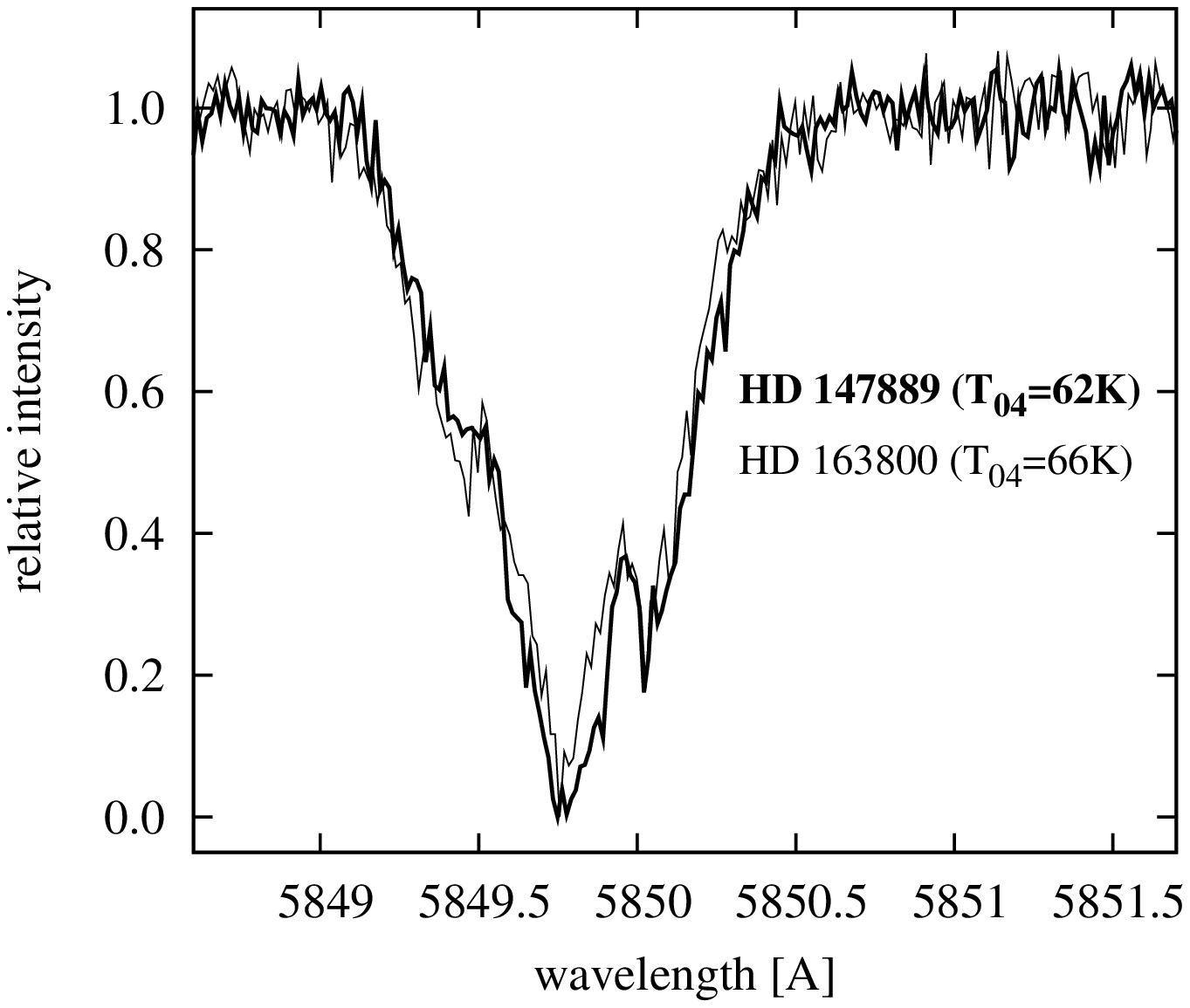}
\includegraphics[width=0.3\textwidth]{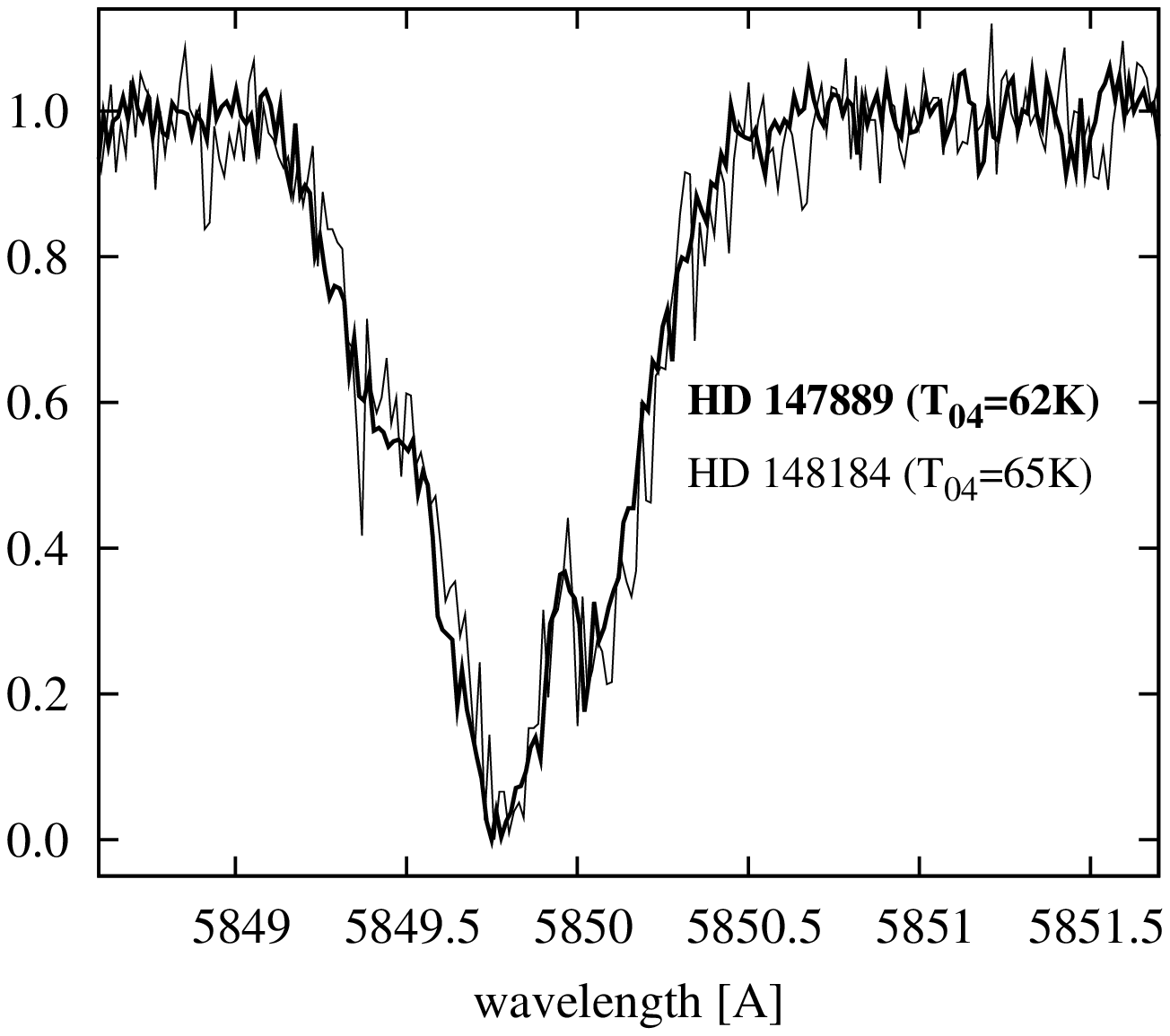}
\includegraphics[width=0.3\textwidth]{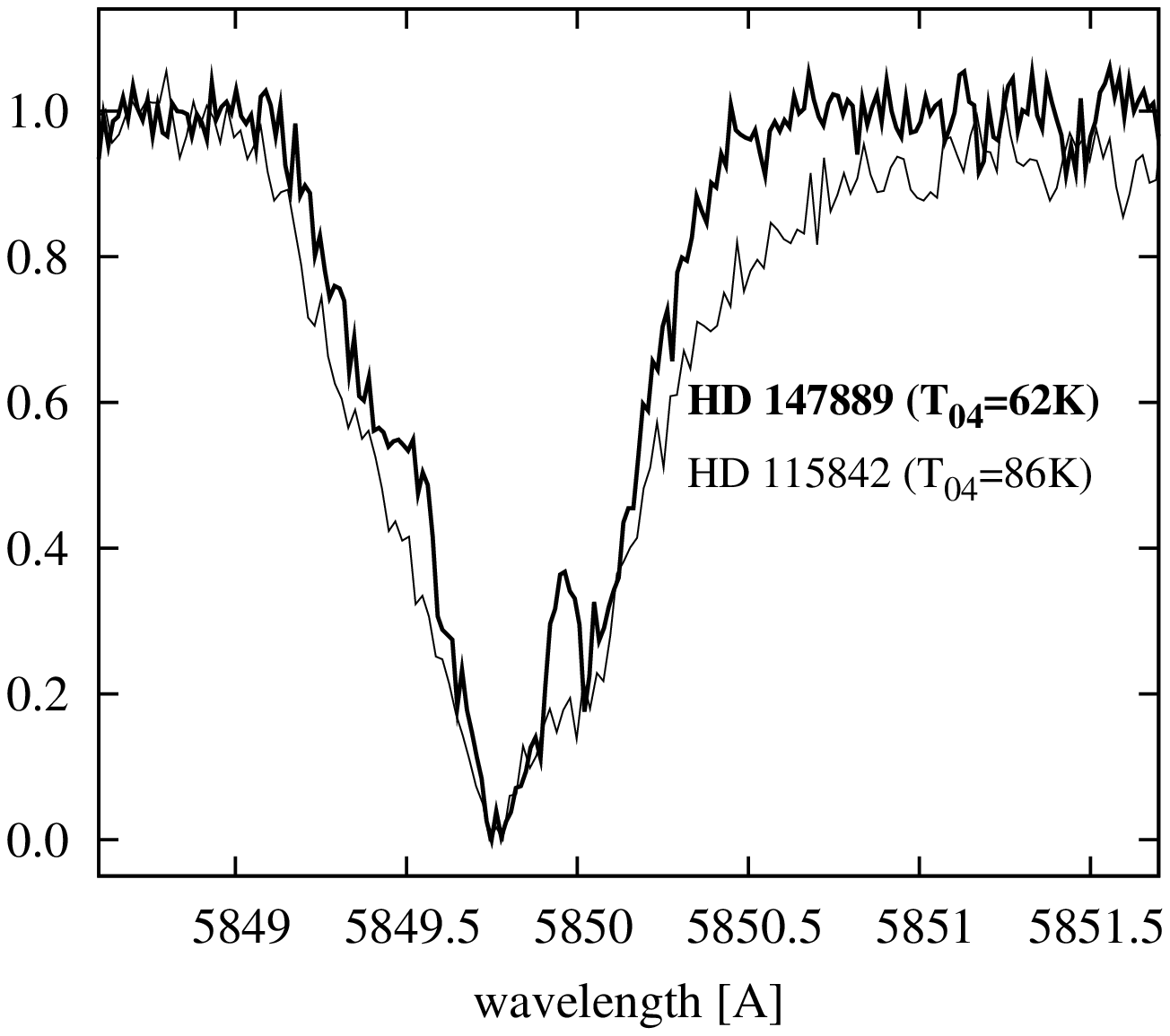}}}
\caption{DIB 5850\,\AA~profiles of the spectra, normalised to their central depths, toward some of the programme stars with different rotational temperatures of $C_2$. DIB profiles does not depend on the excitation temperature of the dicarbon molecule $T_{04}$.}
\label{fig03}
\end{figure*}

\section{Conclusions}

The high resolution and high signal-to-noise ratio spectra acquired using different instruments suggest that for some diffuse interstellar bands exist a tiny effect which can be correlated with rotational temperature of $C_2$. The profile widths of the DIBs at 6196\,\AA~and 5797\,\AA~varies from object to object being broader for higher values of these temperatures. DIBs 4964\,\AA~and 5850\,\AA~do not show that effect. 

\section*{Acknowledgments}

This work was supported by the Science and High Education Ministry of Poland, grants N203 012 32/1550, N203 39/3334 and by UE PhD Scholarship Program (ZPORR).

\bibliographystyle{mn2e}
\bibliography{ms}

\label{lastpage}
\end{document}